\documentclass[a4paper,fleqn]{cas-sc}

\usepackage[authoryear,longnamesfirst]{natbib}

\def\tsc#1{\csdef{#1}{\textsc{\lowercase{#1}}\xspace}}
\tsc{WGM}
\tsc{QE}
\tsc{EP}
\tsc{PMS}
\tsc{BEC}
\tsc{DE}

\usepackage{lineno}

\begin{document}
\let\WriteBookmarks\relax
\def\floatpagepagefraction{1}
\def\textpagefraction{.001}

\title [mode = title]{Combined mechanistic and machine learning method for construction of oil reservoir permeability map consistent with well test measurements}                      
\author[1]{E.A.~Kanin}[orcid=0000-0003-3065-8244]
\cormark[1]
\ead{evgenii.kanin@skoltech.ru}
\author[1]{A.A.~Garipova} 
\author[1]{S.A.~Boronin}
\author[1]{V.V.~Vanovsky}
\author[1]{A.L.~Vainshtein}
\author[2]{A.A.~Afanasyev}
\author[1]{A.A.~Osiptsov}
\author[1]{E.V.~Burnaev}


\address[1]{Skolkovo Institute of Science and Technology (Skoltech), Bolshoy Boulevard 30, bld. 1, Moscow, Russia 121205}
\address[2]{Institute of Mechanics, Moscow State University, Michurinsky Pr., 1, Moscow, Russia 119192}


\begin{abstract}
We propose a new method for construction of the absolute permeability map consistent with the interpreted results of well logging and well test measurements in oil reservoirs. Nadaraya-Watson kernel regression is used to approximate two-dimensional spatial distribution of the rock permeability. Parameters of the kernel regression are tuned by solving the optimization problem in which, for each well placed in an oil reservoir, we minimize the difference between the actual and predicted values of (i) absolute permeability at the well location (results of interpretation of well logging); (ii) absolute integral permeability of the domain around the well and (iii) skin factor (results of interpretation of well tests). Optimization task (inverse problem) is solved via multiple solutions to forward problems, in which we estimate the integral permeability of reservoir surrounding a well and the skin factor by the surrogate model. The last one is developed using an artificial neural network trained on the physics-based synthetic dataset generated using the procedure comprising the numerical simulation of bottomhole pressure decline curve in reservoir simulator followed by its interpretation using a semi-analytical reservoir model. The developed method for reservoir permeability map construction is applied to the available reservoir model (Egg Model) with highly heterogeneous permeability distribution due to the presence of highly-permeable channels. We showed that the constructed permeability map is hydrodynamically similar to the original one. Numerical simulations of production in the reservoir with constructed and original permeability maps are quantitatively similar in terms of the pore pressure and fluid saturations distribution at the end of the simulation period. Moreover, we obtained an good match between the obtained results of numerical simulations in terms of the flow rates and total volumes of produced oil, water and injected water.
\end{abstract}


\maketitle


\section{Introduction}
Building high-quality reservoir simulation models remains a complex task that requires the synergy of several branches of geoscience and reservoir engineering. Traditional approaches to geological model construction, in particular, stochastic modeling, are very time-consuming with the main disadvantage being uncertainty in resulting reservoir properties. For reliable production simulation results, petroleum engineers have to solve the inverse problem, namely, the history matching of the hydrodynamic model. It is an iterative calibration process involving the alteration of the parameters of the original geological model to match the production data. During geological modeling, 2D maps of absolute or effective reservoir permeability are built using static datasets at well locations, which consist of information obtained usually from well logging and core studies (routine and specific core analysis). Accounting for dynamic well test interpretation data is carried out by various techniques but still poses a challenge for engineers. 

The most common way to bring well test data in line with the static data is to adjust the petrophysical relationship between rock porosity and permeability or to tune the variogram used for spatial correlation. These procedures involve manual adjustments according to experience and expertise of a particular engineer. Therefore, the entire process can take a long time and the result can be not optimal in view of hydrodynamic similarity in between the real reservoir property distribution and the constructed permeability and porosity map.

\citet{zakirov2014geostatistically, zakirov2016advanced, zakirov2018geostatistically} presented a highly-promising approach to solve the problem of well logging and well test data fusion in constructing the reservoir permeability map. The authors performed the geostatistically driven history matching by adjoint methods with the set of control parameters including properties of variogram, porosity-permeability relationship for each rock facies, and data at control points. The algorithm is automated and has been successfully validated at several synthetic cases and applied for realistic oilfield models.  

\citet{kolesnikov2010integration} suggested calculating productivity index avoiding reservoir simulations by finite difference approach, which resembles tensor permeability upscaling methods for the rapid calibration of the geological models with well test measurements and production profiles. Modifications of permeability used to calculate productivity indexes allow matching productivity indexes in geological models to the results of well test interpretation. The authors tested the approach on the geological model of an oilfield located in West Siberia. Comparison of actual productivity indexes obtained using well test data with those calculated by the reservoir model after its execution showed an acceptable agreement.

In the study by \citet{he2000conditioning}, authors described a multi-step procedure for the efficient generation of reservoir properties accounting for the dynamic data obtained using stochastic models. Authors claimed that the set of realizations obtained using this algorithm typically provides an acceptable approximation to the probability density function for reservoir models so that the proposed method can be used in Monte Carlo modeling. Authors reported that their approach allowed one to comply with the well test data and retain a heterogeneity typical of the original geological model. It consists of two parts, namely, automatic and manual. The former part is a minimization problem and the latter one is a data preparation task, which requires: (i) cutting a sector from the reservoir model to apply a history matching process; (ii) upscaling the sector model; (iii) determining a variogram using the coarse model, and (iv) choosing the sets of variable and fixed parameters. Even though the automatic part was quite efficient, the procedure could not be utilized as a ``production line'' at that stage.

The method based on the Ensemble Kalman Filter (EnKF) is presented in \citet{coutinho2010conditioning} with the aim to update reservoir permeability distribution using available bottomhole pressure profiles and well logging data, estimate skin factors of reservoir layers and compute ``effective'' skin factor of wells in the framework of multilayered reservoir models. The algorithm worked well for synthetic cases, but it was not successful when applied to the real field case. The authors considered two different approaches: updating layer permeability multipliers with EnKF and double stochastic EnKF, but neither of the two approaches provided a reasonable data match.

Ensemble Kalman Filter was successfully used to perform history matching for real field cases. \citet{evensen2007using} applied this technique to a reservoir located in the North Sea area to estimate permeability, porosity, initial fluid contacts as well as vertical and fault transmissivity multipliers. A similar approach is applied to oil saturated reservoir in the paper \citep{bianco2007history}. The EnKF was also used to conduct history matching for a deepwater formation with multiscale parametrization \citep{zhang2011history}. 

The evolution of EnKF based methods lead to the development of ensemble smoothers (ES) and their use in history matching. In the paper by \citet{evensen2018conditioning}, authors discussed the algorithms for adjusting reservoir models to comply with production rate data by ensemble smoothers. The advantage is that ES allows considering so-called hyperparameters, which represent geological model inputs and less computational expenses compared to EnKF. The authors discussed approaches to reduce redundant information in production data series and how to deal with errors correlated in time. Real field application results have shown that the Iterative ensemble Smoother formulation performed the best in comparison with other formulations.  

As we can see, automated conditioning of dynamic well data is a complex task, and the search for reliable algorithms is still ongoing. Machine learning methods can be applied to significantly speed-up different stages of modeling workflow, for example, seismic and well logging interpretation, facial and petrophysical analysis. Developments in artificial intelligence technologies, especially neural networks, allowed considering the data fusion problems from a different perspective. Several studies were published in the last couple years and dealt with the approaches based on neural networks as described below.

In the study by \citet{thanh2021integrated}, the authors proposed an enhanced framework for modeling the distribution of lithofacies and petrophysical properties of a sandstone reservoir containing fluvial channels. The integrated method is proposed, which is based on (i) artificial neural network (ANN); (ii) Sequential Gaussian Simulation, and (iii) object-based modeling. ANN is used to predict the petrophysical properties of the reservoir by combining seismic attributes and well logging data. Object-based modeling was applied to distribute facies of channels in the 3D model, which allows for resolving realistic depositional environments. The proposed modeling workflow facilitates the reduction of the typical time required to conduct the history matching procedure applied to a reservoir containing fluvial channels.

\citet{bai2020hybrid} suggested the algorithm to construct geologically realistic subsurface models conditioned to well location data with the help of a surrogate algorithm. While cross-correlation-based simulation (CCSIM) allows effective reconstruction of reservoir models, it suffers from the requirement to balance between the quality of realizations and the degree of point data reproduction. The authors combined a pattern-based method with a convolutional neural network (CNN) to overcome this challenge. Inside the surrogate algorithm combining CCSIM and CNN, the former was used for grids in the absence of real data as it allows the generation of high-quality geological features. For the grids, where real data is available, CCSIM is utilized to obtain an initial guess, while CNN is applied to improve initial realizations and increase the accuracy of the real data reproduction. Using the spatial distribution of real data, the proposed method allows one to determine missing domains to cover the mismatched real data. Further, a refill of initial model realizations containing missing zones is carried out using the trained model.

\citet{titus2022conditioning} used ANNs to condition a surface-based geological model (SBGM), constructed with a parametric non-uniform rational B-spline (NURBS) approach to well data. ANNs were applied in the following way: (i) to map input parameters of SGBM to types of facies in the vicinity of well locations in the framework of the forward modeling step and (ii) to obtain the optimized set of input parameters of SBGM using a back-propagation method, so that the constructed SBGM complies with the types of facies obtained in well measurements. The approach was tested on a synthetic 2D case, and it demonstrated the ability to generate a set of realizations that matches the well data. Moving from 2D facies distribution towards petrophysical properties, the authors observed the potential of CNNs and RNNs (recurrent neural networks): the former allows one to learn important features of spatially correlated data, while the latter can be used to condition individual surfaces to model temporal sequences. Moreover, the authors mentioned that the proposed methodology can be reapplied to object-based models, in which a complex reservoir geometry is described by parameterized objects.

As we can see from the literature review, machine learning algorithms has been successfully applied to the problem of production data conditioning and history matching even though it is still a developing field. Currently there is a gap in between existing methods of permeability map construction based on machine learning and well test and well log data fusion. We believe that this important component of geological model construction can be successfully solved provided the corresponding artificial intelligence tool is developed. 

We would like to highlight the importance to construct the absolute permeability field approximation that is hydrodynamically similar to the actual distribution around each well. In the opposite case, when the similarity is absent, one can obtain the significant difference between modeling results and production history when the approximate absolute permeability cube is embedded into a hydrodynamic simulator. Permeability distribution accounting for well test interpretation results can be an optimal initial approximation for the history matching procedure. The main aim of the present work is to develop a computationally efficient algorithm for constructing the absolute permeability cube based on well logging and well test data fusion. We apply numerical and semi-analytical hydrodynamic modeling, and optimization algorithms. Computationally heavy components of this chain of numerical algorithms, namely, reservoir production simulations to obtain pressure decline curve and its consequent interpretation using the analytical reservoir model, are replaced by a fast surrogate model developed using machine learning (ML) algorithm. We demonstrate the capabilities of the proposed combined mechanistic and ML approach using the synthetic case, while the developed method can be used for field data with no modifications. The proposed surrogate algorithm of well test and well log data fusion can be implemented into a wide variety of existing algorithms of permeability map construction to improve the initial guess to overall history matching process by preserving hydrodynamic similarity in between the original and constructed maps.

We organize the paper in the following way. Section \ref{sec:problem_formulation} outlines the problem formulation. Section \ref{sec:modelling_approach} describes the methodology for building the absolute permeability field approximation. In Section \ref{sec:surrogate_model}, we explain the procedure of synthetic dataset generation using the numerical hydrodynamic simulator and semi-analytical reservoir model and describe the surrogate model predicting the integral permeability around the well and skin factor. Section \ref{sec:results} demonstrates the obtained results and their analyses. Section \ref{sec:discussion} provides the main features and limitations of the proposed models for the construction of the absolute permeability map as well as the directions for future development. Finally, we summarize the main findings of our study  and provide potential directions for further research in the area in Section \ref{sec:conclusions}.

\section{Problem formulation}
\label{sec:problem_formulation}
\begin{figure}[pos=htp]
\begin{center}
\centerline{\includegraphics[width=0.8\linewidth]{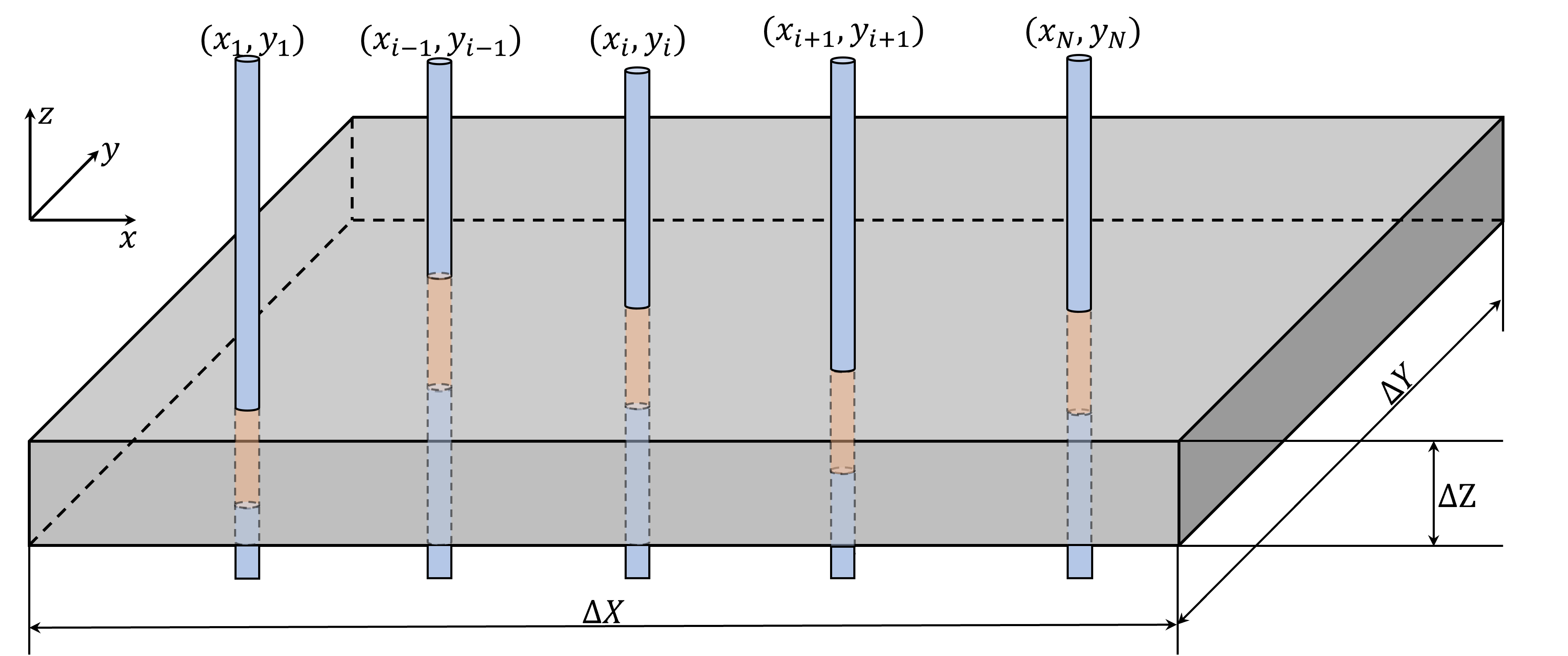}}
\caption{
A synthetic reservoir model; $N$ vertical wells (blue cylinders) cross the formation by perforating it along the entire thickness (red colored zones).
}
\label{fig:reservoir_model}
\end{center}
\end{figure}

The proposed methodology for building the absolute permeability field approximation is outlined in Section \ref{sec:modelling_approach} based on the example of the synthetic reservoir model which we describe in the current Section. Figure \ref{fig:reservoir_model} shows the schematic drawing of the reservoir model. 

Reservoir is approximated by a box with dimensions $\Delta X \times \Delta Y \times \Delta Z$. It is fully penetrated by $N$ vertical wells which are parallel to $z$-axis and located at the points $\left\{x_i, y_i \right\}\big|_{i=1}^N$. We study the formation with homogeneous filtration-storage properties along the vertical direction. As a result, the discussed model is two-dimensional, and our target reservoir property is a function of the lateral coordinates only, $k_0 = k_0(x, y)$, where the subscript ``0'' denotes the real absolute permeability distribution, which is needed to be approximated. 

Assume that the absolute permeabilities ${k^{\mathrm{WL}}_i}$, ${k^{\mathrm{WT}}_i}$ and skin factor $S_i$ are known for each well. Parameter ${k^{\mathrm{WL}}_i}$ corresponds to the absolute permeability at the well location, so that ${k^{\mathrm{WL}}_i} = k_0(x_i, y_i)$ and superscript ``WL'' stands for 'well logging'. Using the well logging measurements, one can determine the rock porosity $\phi$ in the vicinity of the well, which can be converted into absolute permeability via the dependence $k^{\mathrm{WL}}(\phi)$ obtained using lab experiments on rock core samples. Parameter ${k^{\mathrm{WT}}_i}$ is an integral permeability of the rock surrounding the the well, so that
\begin{equation}
\label{eq:F}
{k^{\mathrm{WT}}_i} = \mathcal{F}(k_0(x, y)), ~ (x, y) \in \mathcal{A}^{\mathcal{R}}_i,    
\end{equation}
where $\mathcal{A}^{\mathcal{R}}_i$ is a circle round the well $i$ of radius $\mathcal{R}$, and superscript ``WT'' denotes `well test'. By function $\mathcal{F}$ we denote the physics-based averaging of the permeability field around the well inside the circle of radius $\mathcal{R}$, which can be interpreted as a distance from the well up to which the pore pressure disturbance propagate during the well test. Skin factor $S_i$ quantifying the contrast between the rock permeability in the very vicinity of the well and at a certain distance from it, is also a function of the absolute permeability field around the well 
\begin{equation}
\label{eq:G}
S_i = \mathcal{G}(k_0(x, y)), ~ (x, y) \in \mathcal{A}^{\mathcal{R}}_i.
\end{equation}

The absolute integral permeability ${k^{\mathrm{WT}}_i}$ and skin factor $S_i$ can be found from the well test analysis. The standard method is interpretation of the bottomhole pressure build-up curve during the well shut-in (build-up test) \citep{horne1995modern}. Using the approach proposed by \citet{perrine1956analysis} and \citet{martin1959simplified} including the concepts of total mobility and total compressibility, one can estimate the total mobility $\lambda_t$ and skin factor values from the interpretation. Further, absolute permeability is computed using the equation

\begin{equation*}
    k^{\mathrm{WT}} = \lambda_t\left(\frac{k_{ro}}{\mu_o} + \frac{k_{rw}}{\mu_w} + \frac{k_{rg}}{\mu_g}\right)^{-1},
\end{equation*}
where $k_{ro}$, $\mu_{o}$, $k_{rw}$, $\mu_{w}$, $k_{rg}$, $\mu_{g}$,  are relative permeabilities and viscosities of oil, water, and gas phase, respectively. Saturations of each phase in the reservoir are required to compute $k_{ro}, k_{rw}, k_{rg}$. Since saturations depend on time and distance from the well during test, and there are certain difficulties in their measurement, $k_{\mathrm{WT}}$ value is usually determined with significant error. 

In the current work, we construct the approximation $k = k(x, y)$ of the real absolute permeability field $k_0 = k_0(x, y)$ assuming that the well locations $\left\{x_i, y_i \right\}\big|_{i=1}^N$, permeability values obtained from the well logging $\left\{k^{\mathrm{WL}}_i\right\}\big|_{i=1}^N$, integral permeabilities $\left\{k^{\mathrm{WT}}_i\right\}\big|_{i=1}^N$, and skin factor values $\left\{S_i\right\}\big|_{i=1}^N$ are available. Since we account for the integral permeability in the approximation, the absolute permeability map around each well is hydrodynamically similar to the real distribution, which allows to obtain acceptable match between the results of reservoir simulations and observed production data already at the start of history matching process. 

\section{Global model for reservoir permeability map construction}
\label{sec:modelling_approach}
In the framework of current study we approximate the absolute permeability map of reservoir $k_0 = k_0(x, y)$ using Nadaraya-Watson kernel regression \citep{nadaraya1964estimating, watson1964smooth}. According to this approach, functional dependence of absolute permeability on the spatial coordinates has the following form:
\begin{equation}
    k(x, y) = \frac{\sum_{i=1}^N\left[k^{\mathrm{near}}_i\mathcal{K}^{\mathrm{near}}(r-r_i) + k^{\mathrm{far}}_i\mathcal{K}^{\mathrm{far}}(r-r_i)\right]}{\sum_{i=1}^N\left[\mathcal{K}^{\mathrm{near}}(r-r_i) + \mathcal{K}^{\mathrm{far}}(r-r_i)\right]},
    \label{eq:kernel_regression}
\end{equation}
where $r = \sqrt{x^2 + y^2}$, $r_i = \sqrt{x_i^2 + y_i^2}$ and kernel functions:
\begin{equation}
    \mathcal{K}^{\mathrm{far}}(r) = \left(\frac{r}{r_d}\right)^{\alpha} \mathrm{exp}\left[-\left(\frac{r}{r_d}\right)^\beta\right], ~~~~ \mathcal{K}^{\mathrm{near}}(r) = \gamma ~ \mathrm{exp}\left[-\left(\frac{r}{r_g}\right)^\delta\right]
    \label{eq:kernels}
\end{equation}

Equations \eqref{eq:kernel_regression} and \eqref{eq:kernels} contain $2N + 6$ unknown parameters: 
\begin{equation*}
    \Phi = \left\{\Phi_i\right\}\big|_{i=1}^N = \left\{k^{\mathrm{near}}_i, k^{\mathrm{far}}_i\right\}\big|_{i=1}^N,\quad
\Psi = \left\{r_d, r_g, \alpha, \beta, \gamma, \delta \right\},
\end{equation*} 
which we denote by $\Omega = \{\Phi, \Psi\}$. The former ones, $\Phi$, have the dimension of permeability (in millidarcy range in the framework of current study), and describe the contribution of permeabilities in the near and far zones of each well to the reservoir absolute permeability map $k(x, y)$ calculated using Eq.~\eqref{eq:kernel_regression}. The contributions are weighted by the kernel functions \eqref{eq:kernels} parameterized by characteristic lengthscales $r_d$ and $r_g$, which have the dimension of length, as well as dimensionless variables $\alpha, \beta, \gamma, \delta$, which are the fixed and assumed to be similar of all wells. A typical spacial distribution of kernel functions \eqref{eq:kernel_regression_2} is shown in  Fig.~\ref{fig:reservoir_model}. One can observe that function $\mathcal{K}^{\mathrm{near}}(r)$ takes maximum value at $r = 0$, so that its the main contribution to overall permeability map $k(x,\,y)$ is at the point of the well location $(x_i, y_i)$. At the same time, $\mathcal{K}^{\mathrm{far}}(r)$ reaches the maximum at the distance $r = (\alpha/\beta)^{1/\beta}r_d$, which describes the far-field contribution to absolute permeability map according to well test data at the distance $r_d$ scaled by the combination of parameters $\alpha$ and $\beta$.      

\begin{figure}[pos=htp]
\begin{center}
\centerline{\includegraphics[width=0.6\linewidth]{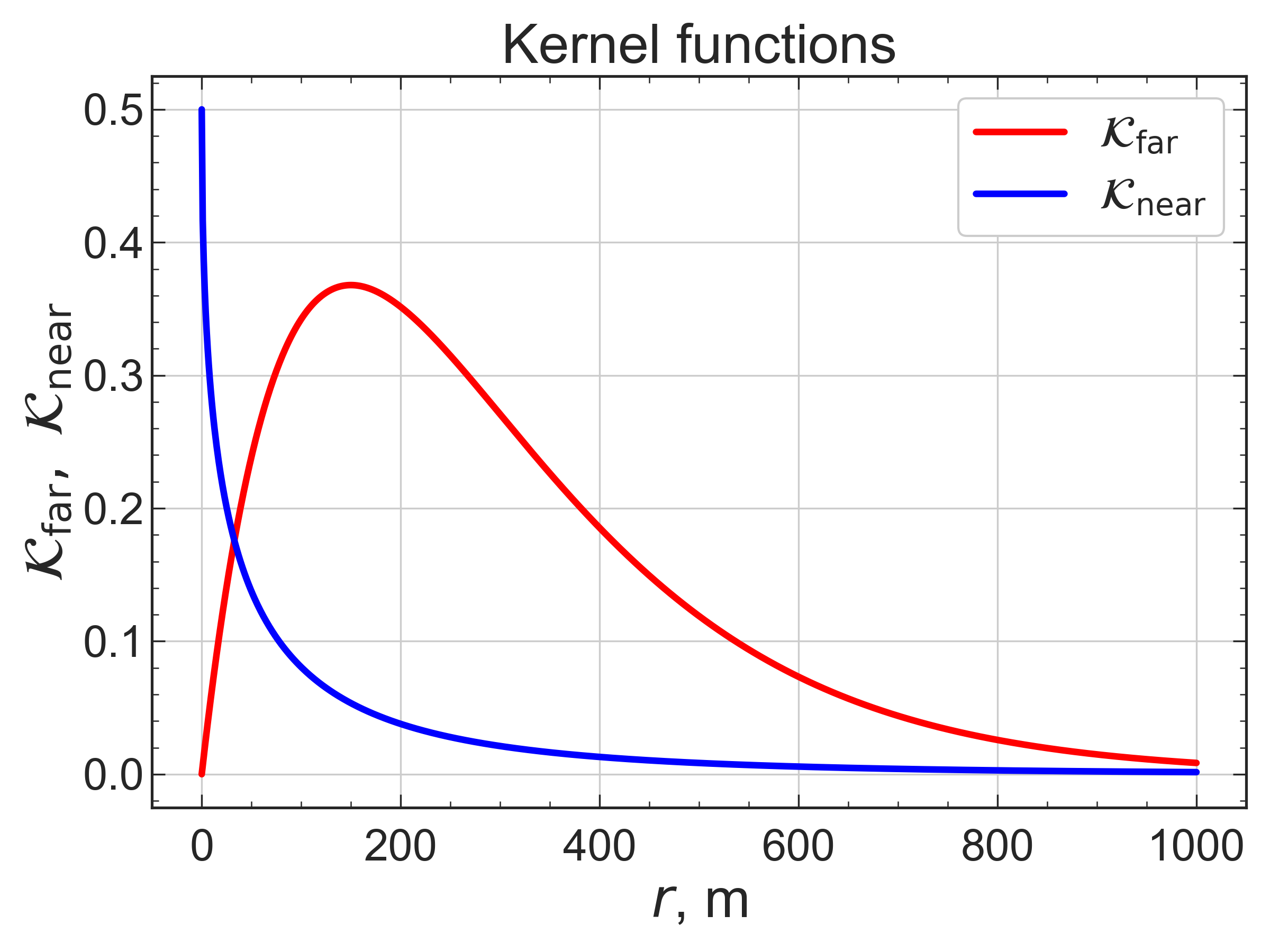}}
\caption{
Kernel functions $\mathcal{K}^{\mathrm{far}}(r)$ and $\mathcal{K}^{\mathrm{near}}(r)$ (see Eq.~\eqref{eq:kernel_regression_2}) at $r_d = 150 ~ \mathrm{m}$, $r_g = 30 ~ \mathrm{m}$, $\alpha = 1, ~ \beta = 1, ~ \gamma = 0.5, ~ \delta = 0.5$.
}
\label{fig:kernels}
\end{center}
\end{figure}

Parameters $\Omega$ are calculated using the available formation properties described in Section \ref{sec:problem_formulation}: $\left\{k^{\mathrm{WL}}_i, k^{\mathrm{WT}}_i, S_i\right\}\big|_{i=1}^N$. They are obtained via solution of the minimization problem:
\begin{equation}
    \min_{\Omega} \frac{1}{N}\sum_{i=1}^N \big|\big| \widetilde{\mathcal{X}}^*_i - \mathcal{X}^*_i \big|\big|_1,
    \label{eq:minimization_1}
\end{equation}
where $||\cdot ||_1$ denotes the $L_1$-norm. Vector $\mathcal{X}_i$ is composed of the ``true'' values of absolute permeabilities obtained via well logging and well tests as well as skin factors: $\mathcal{X}_i = \left\{k^{\mathrm{WL}}_i, k^{\mathrm{WT}}_i, S_i\right\}$; vector $\widetilde{\mathcal{X}}$ is composed of similar parameters estimated using the approximate absolute permeability field \eqref{eq:kernel_regression}, $\widetilde{\mathcal{X}}_i = \left\{k(x_i, y_i), \widetilde{k}_i, \widetilde{S}_i\right\}$; the superscript ``*'' denotes that the parameters are scaled before substituting into target function \eqref{eq:minimization_1}. We apply standard scaling, so that vector has zero mean and unit variance after the transformation with the parameters determined from the normalization of matrix $\mathcal{X} = \left\{\mathcal{X}_i\right\}\big|_{i=1}^N$. We solve the minimization problem \eqref{eq:minimization_1} using differential evolution optimization algorithm implemented in SciPy library \citep{2020SciPy-NMeth}.

We predict the values of integral permeability $\{\widetilde{k}_i\}_{i=1}^N$ and skin factor $\{\widetilde{S}_i\}_{i=1}^N$ for each well using the surrogate model. In fact, this model approximates functions $\mathcal{F}$ and $\mathcal{G}$ introduced in Section \ref{sec:problem_formulation} (see Eqs.~\eqref{eq:F}, \eqref{eq:G}). Details of the development of the surrogate model are formulated below in Section~\ref{sec:surrogate_model}, while in the rest of this section we discuss application of the developed model to the construction of absolute permeability map. 

Let us consider the well denoted by index $j$. We clip a square domain of size $\mathcal{D}$ with sides parallel to the axes $x$ and $y$, which is formally described as 
\begin{equation*}
    \mathcal{B}^{\mathcal{D}}_j = \left[x_j - \frac{\mathcal{D}}{2}, x_j + \frac{\mathcal{D}}{2}\right] \times  \left[y_j - \frac{\mathcal{D}}{2}, y_j + \frac{\mathcal{D}}{2}\right].
\end{equation*}

The absolute permeability distribution is given by equation \eqref{eq:kernel_regression}, which we rewrite in the following way
\begin{flalign}
    & k(x, y) = \frac{\mathcal{M}_j + \sum_{i=1, i\neq j}^N \mathcal{M}_i}{\mathcal{N}_j + \sum_{i=1, i\neq j}^N \mathcal{N}_i}, \nonumber \\ 
    & \mathcal{M}_i = k^{\mathrm{near}}_i\mathcal{K}^{\mathrm{near}}(r-r_i) + k^{\mathrm{far}}_i\mathcal{K}^{\mathrm{far}}(r-r_i), \nonumber \\ 
    & \mathcal{N}_i = \mathcal{K}^{\mathrm{near}}(r-r_i) + \mathcal{K}^{\mathrm{far}}(r-r_i).
    \label{eq:kernel_regression_1}
\end{flalign}
Assuming that the distance between the wells is large enough, the contribution from well $j$ into the absolute permeability field $k(x, y)$ is given by the ratio $k^{\mathrm{well}}_j = \mathcal{M}_j/\mathcal{N}_j$, which is described by the set of parameters $\{\Phi_j, \Psi\}$. In turn, the impact of the neighbouring wells on the permeability distribution $k(x, y)$ is governed by the function $k^{\mathrm{neigh}}_j = \sum_{i=1, i\neq j}^N \mathcal{M}_i / \sum_{i=1, i\neq j}^N \mathcal{N}_i$. One can approximate the function $k^{\mathrm{neigh}}_j$ inside the zone $\mathcal{B}^{\mathcal{D}}_j$ by a quadratic polynomial with two variables:
\begin{flalign}
    & \widetilde{k}^{\mathrm{neigh}}_j = a_{2, 0} (x - x_j)^2 + a_{1, 1} (x-x_j) (y-y_j) + a_{0, 2} (y-y_j)^2 + \nonumber \\
    & + a_{1, 0} (x-x_j) + a_{0, 1} (y-y_j) + a_{0, 0}. 
    \label{eq:quadratic_polynomial}
\end{flalign}
Consequently, there is a one-to-one correspondence between the permeability distribution around the well $j$ described as $k(x, y), ~ (x, y) \in \mathcal{B}^{\mathcal{D}}_j$, and the set of 14 parameters:
\begin{equation}
    \chi = \left\{\underbrace{k^{\mathrm{near}}_j, k^{\mathrm{far}}_j,}_{\Phi_j} \underbrace{r_d, r_g, \alpha, \beta, \gamma, \delta,}_{\Psi} \underbrace{a_{2, 0}, a_{1, 1}, a_{0, 2}, a_{1, 0}, a_{0, 1}, a_{0, 0}}_{\Xi_j} \right\},
    \label{eq:params_ml_model}
\end{equation}
where we introduce notation $\Xi_j$ for the quadratic polynomial coefficients. Based on the values of the parameters \eqref{eq:params_ml_model}, our surrogate model (Section~ \ref{sec:surrogate_model}) estimates the integral permeability $\widetilde{k}_j$ and the skin factor $\widetilde{S}_j$. 

\section{Surrogate model for evaluation of integral permeability}
\label{sec:surrogate_model}
In this section, we describe the surrogate model predicting the integral permeability around the well (mimicking the results of well test interpretation) and corresponding skin factor. The model is based on machine learning algorithm, namely, artificial neural network (ANN) \citep{rosenblatt1958perceptron}. ANN is trained on the physics-based synthetic dataset generated with the help of the numerical hydrodynamic simulator MUFITS \citep{afanasyev2020mufits} and in-house semi-analytical reservoir model \citep{ozkan1991new}.

We begin with the explanation of the synthetic dataset preparation procedure. We create multiple instances of synthetic reservoir model similar to that described in Section~\ref{sec:problem_formulation} (Figure \ref{fig:reservoir_model}), namely, a rectangular box with dimensions $\Delta X = \Delta Y = 4 ~ \mathrm{km}$, $\Delta Z = 10 ~ \mathrm{m}$ (the thickness value is for reference only since the model is effectively two-dimensional). 

\begin{figure}[pos=htp]
\begin{center}
\centerline{\includegraphics[width=0.6\linewidth]{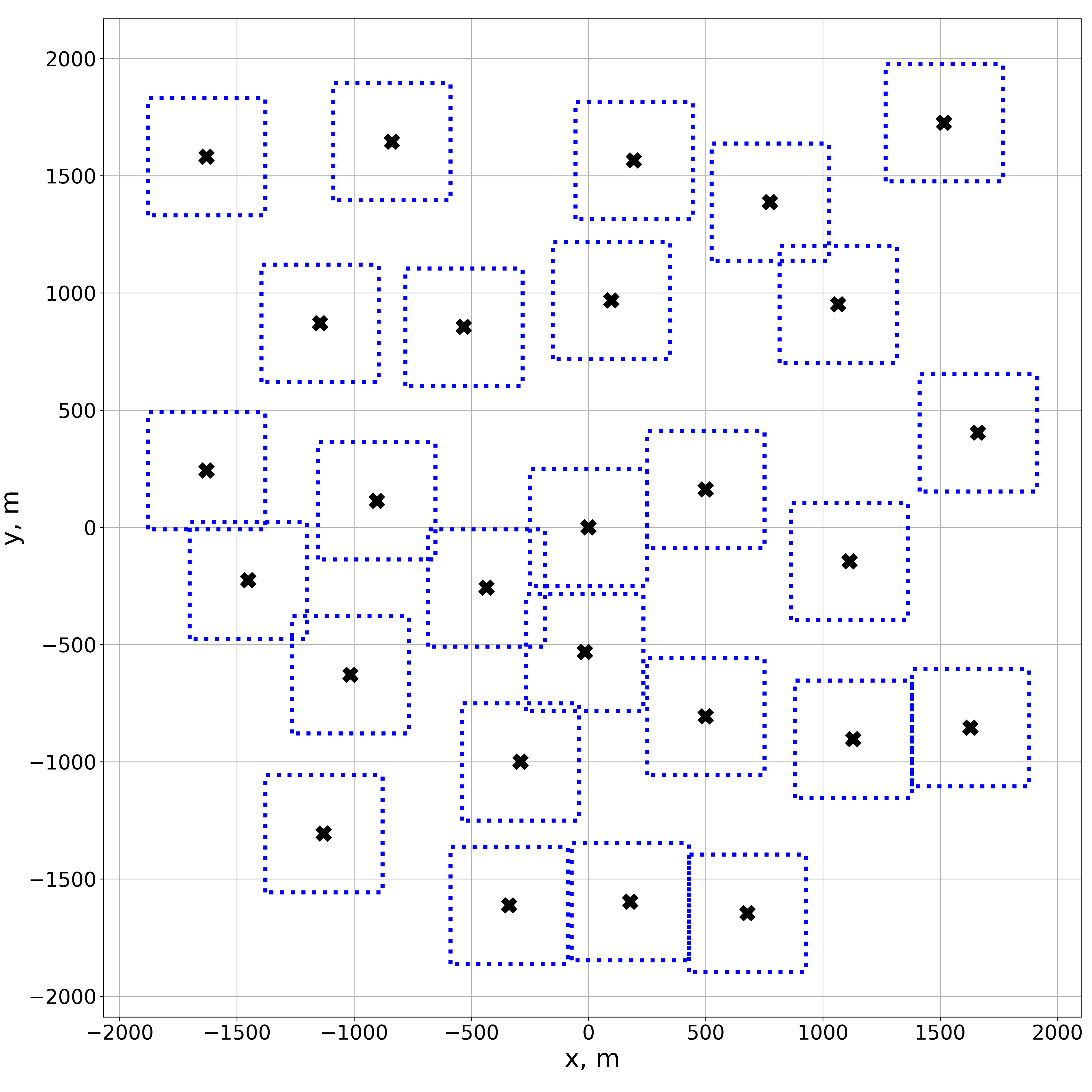}}
\caption{
An example of the synthetic reservoir model generated according to the algorithm described in the main text. Here, the minimal distance between any two wells $\Delta d$ equals 500 m. Black crosses denote the positions of the vertical wells, while the squares with dotted blue boundary mark the domains $\mathcal{B}^{\mathcal{D}}$ used for the creation of the samples for the synthetic dataset.    
}
\label{fig:syntheric_formation_for_database}
\end{center}
\end{figure}

Let us consider the generation of a single realization of the synthetic reservoir model. Vertical wells are placed randomly subject to the condition that the distance between any two wells is greater then the predefined value $\Delta d$, so that $\sqrt{(x_i - x_j)^2 + (y_i - y_j)^2} \geq \Delta d$ for any $i \ne j$. For a particular reservoir model, distance $\Delta d$ is chosen randomly in the range $\Delta d \in [300, 600] ~ \mathrm{m}$. The placement of vertical wells continues until the algorithm can not find any possible location for the next well. Figure~\ref{fig:syntheric_formation_for_database} illustrates the results of application of the described procedure at $\Delta d$ = 500 m. 

At the next step, the algorithm assigns the set of parameters for each well $\left\{\Phi_i, \Psi_i\right\}_{i=1}^N$, where $\Psi_i$ is the set of kernel parameters $\{r_d, r_g, \alpha, \beta, \gamma, \delta\}$ specified for each of the wells. This feature of the absolute permeability map parametrization differs from the one utilized in the general modelling approach (Section~\ref{sec:modelling_approach}) allowing us to use diversified samples from a single synthetic reservoir model into the overall dataset. The values of parameters $\left\{\Phi_i, \Psi_i\right\}_{i=1}^N$ are selected randomly in the following ranges:
\begin{flalign}
    & k^{\mathrm{near}}, k^{\mathrm{far}} \in [1, 15] ~ \mathrm{mD}, ~ r_d \in [100, 300] ~ \mathrm{m}, ~ r_g \in [5, 50] ~ \mathrm{m}, \nonumber \\
    & \alpha \in [0.5, 2], ~ \beta \in [1, 2], ~ \gamma \in [0.01, 2], ~ \delta \in [0.05, 1].
    \label{eq:value_ranges}
\end{flalign}
The range for variables $k^{\mathrm{near}}, k^{\mathrm{far}}$ are chosen in accordance with the typical permeability values of an oilfield in Western Siberia, while the intervals for the geometrical parameters $r_d, r_g$ are chosen according to the typical areas around wells covered by well logging and well tests. We select experimentally the ranges for the remaining parameters of the kernel regression, namely, $\alpha, \beta, \gamma, \delta$.

Next, we apply the modified version of kernel regression \eqref{eq:kernel_regression} to compute the absolute permeability field:
\begin{flalign}
    & k(x, y) = \frac{\sum_{i=1}^N\left[k^{\mathrm{near}}_i\mathcal{K}^{\mathrm{near}}_i(r-r_i) + k^{\mathrm{far}}_i\mathcal{K}^{\mathrm{far}}_i(r-r_i)\right]}{\sum_{i=1}^N\left[\mathcal{K}^{\mathrm{near}}_i(r-r_i) + \mathcal{K}^{\mathrm{far}}_i(r-r_i)\right]}, \nonumber \\
    & \mathcal{K}^{\mathrm{far}}_i(r) = \left(\frac{r}{r_{d,i}}\right)^{\alpha_i} \mathrm{exp}\left[-\left(\frac{r}{r_{d,i}}\right)^{\beta_i}\right], \nonumber \\
    & \mathcal{K}^{\mathrm{near}}_i(r) = \gamma_i ~ \mathrm{exp}\left[-\left(\frac{r}{r_{g,i}}\right)^{\delta_i}\right ]
    \label{eq:kernel_regression_2}
\end{flalign}

\begin{figure}[pos=htp]
\begin{center}
\centerline{\includegraphics[width=1\linewidth]{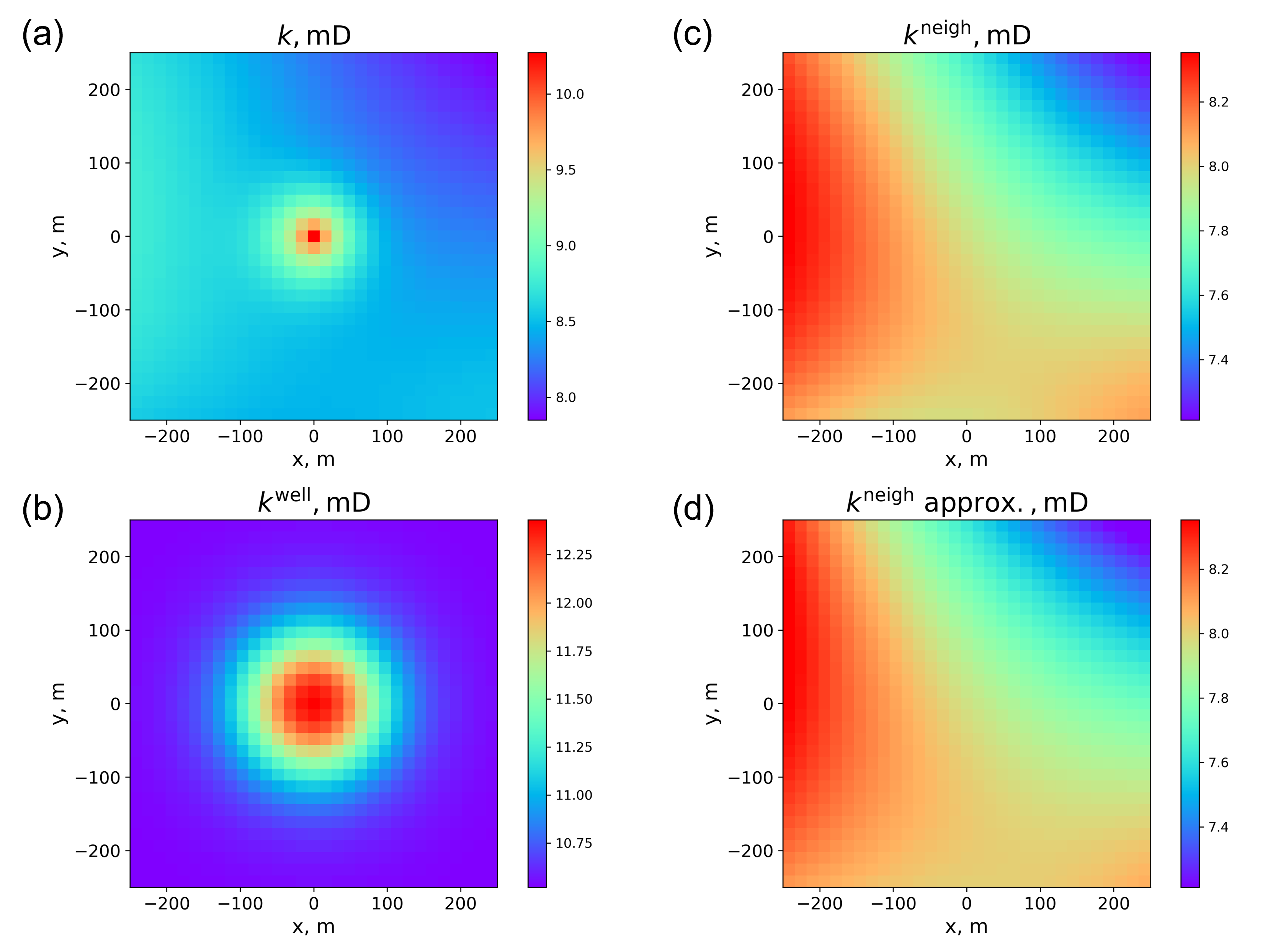}}
\caption{
Plot (a) shows an example of the synthetic permeability distribution $k(x, y)$ in the square domain $\mathcal{B}^{\mathcal{D}}$; contributions from well $k^{\mathrm{well}}$ located in the center of the domain and neighbor wells $k^{\mathrm{neigh}}$ are shown in plots (b) and (c); distribution $k^{\mathrm{neigh}}$ is fitted by the quadratic polynomial  \eqref{eq:quadratic_polynomial}, and plot (d) shows the result of approximation.    
}
\label{fig:database_sample}
\end{center}
\end{figure}

After that, the algorithm is applied to each well and performs the following operations. It cuts a square of size $\mathcal{D} = 500 ~ \mathrm{m}$ with sides parallel to the coordinate axes (domain $\mathcal{B}_j^{\mathcal{D}}$). The algorithm identifies the contributions from well $j$ ($k_j^{\mathrm{well}}$, see Figure~\ref{fig:database_sample}b) and neighbor wells ($k_j^{\mathrm{neigh}}$, see Figure~\ref{fig:database_sample}c) into the absolute permeability field $k(x, y), (x, y) \in \mathcal{B}_j^{\mathcal{D}}$ (see Figure~\ref{fig:database_sample}a). The former distribution is characterized by parameters $\{\Phi_j, \Psi_j\}$, while the latter one is governed by the quadratic polynomial coefficients  \eqref{eq:quadratic_polynomial}; the approximation of $k_j^{\mathrm{neigh}}$ by the polynomial is shown in Figure~\ref{fig:database_sample}d.

Finally, we save the absolute permeability distribution $k(x, y), (x, y) \in \mathcal{B}_j^{\mathcal{D}}$ into the text file for consequent numerical simulations in MUFITS, while the vector $\chi = \left\{ \Phi_j, \Phi_j, \Xi_j \right\}$ is stored into the table of input parameters. This completes the input features preparation procedure for the surrogate model. 

Now, we discuss the estimation of the output parameters for neural network, namely, integral permeability and skin factor. The procedure can be called as the synthetic well test, and it consists of two stages: 
\begin{enumerate}
    \item the numerical simulations of a drawdown test: bottomhole pressure dynamics $p^{\mathrm{num}}_w$ is simulated using MUFITS provided the fixed well flow rate;
    \item interpretation of the bottomhole pressure dynamics using the semi-analytical reservoir model; we minimize the $L_2$-norm of the vector composed of the deviations between the numerical and semi-analytical $p^{\mathrm{s-a}}_w(t)$ pressure values in different time instants within the specified interval:
    \begin{equation}
        \min_{\widetilde{k}, \widetilde{S}} \big|\big|\mathcal{P}\big|\big|_2,
        \label{eq:minimization_2}
    \end{equation}
    where $\mathcal{P} = \{p^{\mathrm{num}}_w(t_i) - p^{\mathrm{s-a}}_w(t_i)\}_{i=1}^T$. The minimization problem is solved using the gradient method Nelder–Mead implemented in SciPy library \citep{2020SciPy-NMeth}.
\end{enumerate}

We specify the following input parameters in the numerical and semi-analytical hydrodynamic models: 
\begin{itemize}
    \item formation dimensions are $\Delta x = \Delta y = 500$ m, $\Delta z = 10$ m;
    \item well radius is $r_w = 0.1$ m;
    \item porosity is $\phi = 15~\%$;
    \item fluid parameters: viscosity is $\mu = 2.5$ cP, total compressibility is $c_t = 2 \cdot 10^{-4}$ bar$^{-1}$, formation volume factor is $B = 1.2$ m$^3$/sm$^3$;
    \item initial condition is uniform pore pressure $p_i = 250$ bar;
    \item boundary conditions are constant pressure $p = p_i$ at lateral borders, top and bottom boundaries are closed;
    \item flow rate is $q = 20$ m$^3$/d;
    \item production period is $t \in [0, 180]$ d.
\end{itemize}

Each permeability distribution $k(x, y), (x, y) \in \mathcal{B}^{\mathcal{D}}$ obtained using Eq.~\eqref{eq:kernel_regression_2} and stored in a text file is passed to numerical hydrodynamic simulator. In the semi-analytical hydrodynamic model, permeability $\widetilde{k}$ is uniform and its value, as well as the skin factor $\widetilde{S}$, are determined by the solution of the minimization problem \eqref{eq:minimization_2}. Consequently, $\widetilde{k}$ parameter is assumed to be the integral permeability corresponding to the distribution $k(x, y)$ inside the domain $\mathcal{B}^{\mathcal{D}}$, and the adjusted homogeneous permeability distribution is hydrodynamically similar to the heterogeneous absolute permeability field $k(x, y)$. Note that the described physics-based averaging procedure for the absolute permeability field does not depend on fluid and rock properties, boundary conditions and operation mode since the absolute permeability is a geometrical parameter of the rock. Integral permeability is determined according to the bottomhole pressure dynamics during the transient production period at the interpretation stage.

In the numerical reservoir model, approximation is carried out using the mesh with cell size of 15 m $\times$ 15 m. Local grid refinement is applied in the vicinity of the wellbore with the cell size of 5 m. Note the the choice of mesh resolution is based on numerical convergence tests. Filtration in the reservoir is simulated using BLACKOIL module, and we enable the single-phase fluid option, namely, a dead oil.

Semi-analytical reservoir model is represented by a fully-penetrating vertical line-source well in a reservoir, the analytical solution is carried out in the Laplace space. The inverse Laplace transformation is performed using Stehfest numerical algorithm \citep{stehfest1970algorithm}. To give some details on the analytical solution, it is derived using the principle of superposition according to which, the solution for a uniform-flux point source along the well trajectory is integrated \citep{ozkan1988performance, ozkan1991new}. The latter function is an analytical solution to 3D filtration equation for slightly compressible fluid in the reservoir approximated by a rectangular box with homogeneous properties, in which the uniform-flux point source is located. The mathematical formulation of the point-source problem is as follows \citep{van1949application, hovanessian1961pressure}: 
\begin{flalign*}
    & \frac{\widetilde{k}}{\mu}\left[\frac{\partial^2 p}{\partial x^2}+\frac{\partial^2 p}{\partial y^2}+\frac{\partial^2 p}{\partial z^2}\right] = \phi c_t \frac{\partial p}{\partial t} + \frac{qB}{\Delta z} \delta(x-x_w)\delta(y-y_w)\delta(z-z_w), \\
    & (x, y, z) \in [0, \Delta x] \times [0, \Delta y] \times [0, \Delta z];  \\
    & p(t=0) = p_i; \\
    & p(x=0) = p(x=\Delta x) = p(y=0)  
    = p(y=\Delta y) = p(z=0) = p(z=\Delta z) = p_i;
\end{flalign*}
where coordinates $(x_w, y_w, z_w)$ is the point source location and $\delta(\cdot)$ is the Dirac delta function. Bottomhole pressure dynamics $p^{\mathrm{s-a}}_w(t)$ corresponds to the pore pressure evolution at the point $x = r_w$, $y = 0$. The analytical solutions for a uniform-flux point source and uniform-flux fully-penetrating vertical line-source well can be found in papers by \citet{ozkan1988performance, ozkan1991new}. We apply several techniques improving the convergence of the series and shortening the computation time as described in \citet{ozkan1988performance, ozkan1991new_2, ozkan1994new}.   

\begin{figure}[pos=htp]
\begin{center}
\centerline{\includegraphics[width=1\linewidth]{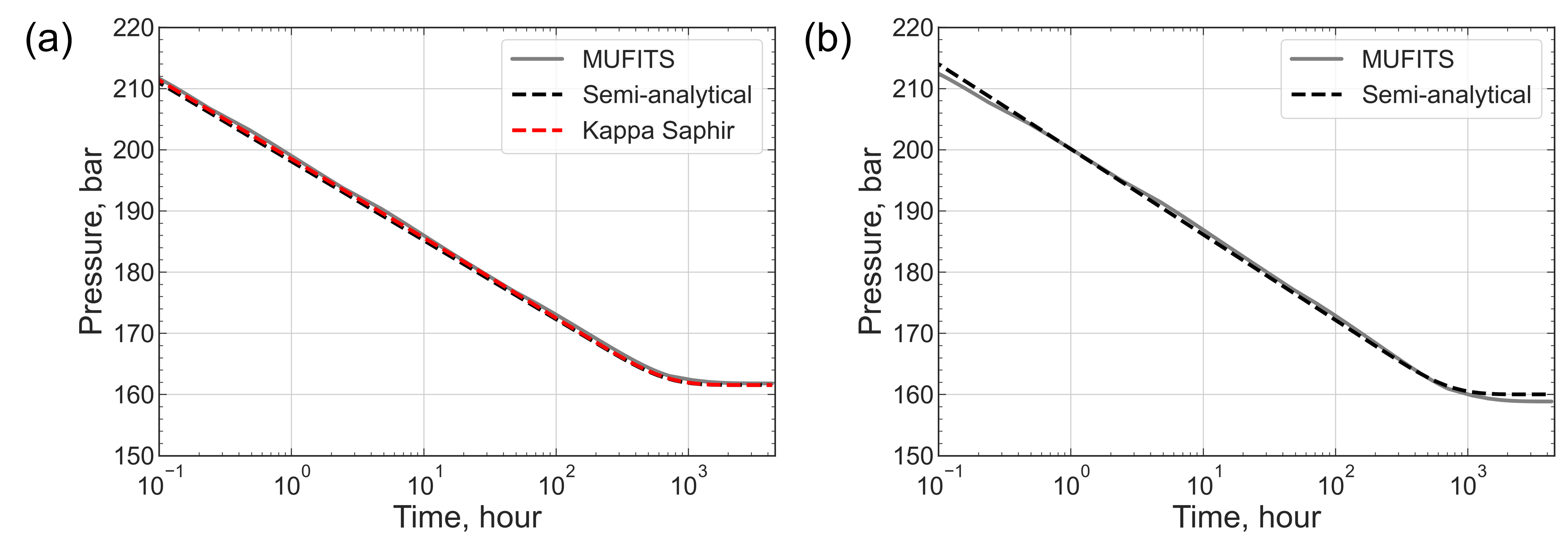}}
\caption{
Plot (a) shows comparison of bottomhole pressure dynamics obtained in in-house semi-analytical reservoir model (dashed black curve), numerical model in MUFITS simulator (solid gray curve) and semi-analytical model implemented into Kappa Saphir (dashed red curve); in plot (b) we show the results of solution to minimization problem \eqref{eq:minimization_2} (welltest interpretation using semi-analytical model) in a reservoir with the non-uniform absolute permeability distribution shown in Fig.~\ref{fig:database_sample}a.  
}
\label{fig:verification_and_application_example}
\end{center}
\end{figure}

Verification of the developed semi-analytical reservoir filtration model is carried out (see Fig.~\ref{fig:verification_and_application_example}a). For this purpose we consider the formation with uniform permeability of 10 mD and compare the bottomhole pressure behavior as a function of time computed via the results obtained  using numerical $p^{\mathrm{num}}_w(t)$ and semi-analytical $p^{\mathrm{s-a}}_w(t)$ models as implemented into MUFITS simulator and commercial software Kappa Saphir, respectively. We obtain a good match between Kappa Saphir and in-house semi-analytical model, while there is a small discrepancy between the results of simulations conducted in MUFITS simulator and benchmark solution in Kappa Saphir. The latter one can be attributed to implementation of the equation of state describing slightly compressible fluid embedded into BLACKOIL module of MUFITS simulator, where the density dependence on pressure includes both linear the quadratic terms. In Figure \ref{fig:verification_and_application_example}b, we show the results of interpretation of welltest in the reservoir with the permeability distribution $k(x, y)$ presented in Figure \ref{fig:database_sample}a. Here, we compare the time dependencies $p^{\mathrm{num}}_w(t)$ and $p^{\mathrm{s-a}}_w(t)$ computed via the numerical simulator MUFITS and semi-analytical reservoir model, respectively. The latter curve corresponds to the integral permeability 9.22 mD and skin factor -0.49 obtained by solving the minimization problem \eqref{eq:minimization_2}. 

\begin{figure}[pos=htp]
\begin{center}
\centerline{\includegraphics[width=0.6\linewidth]{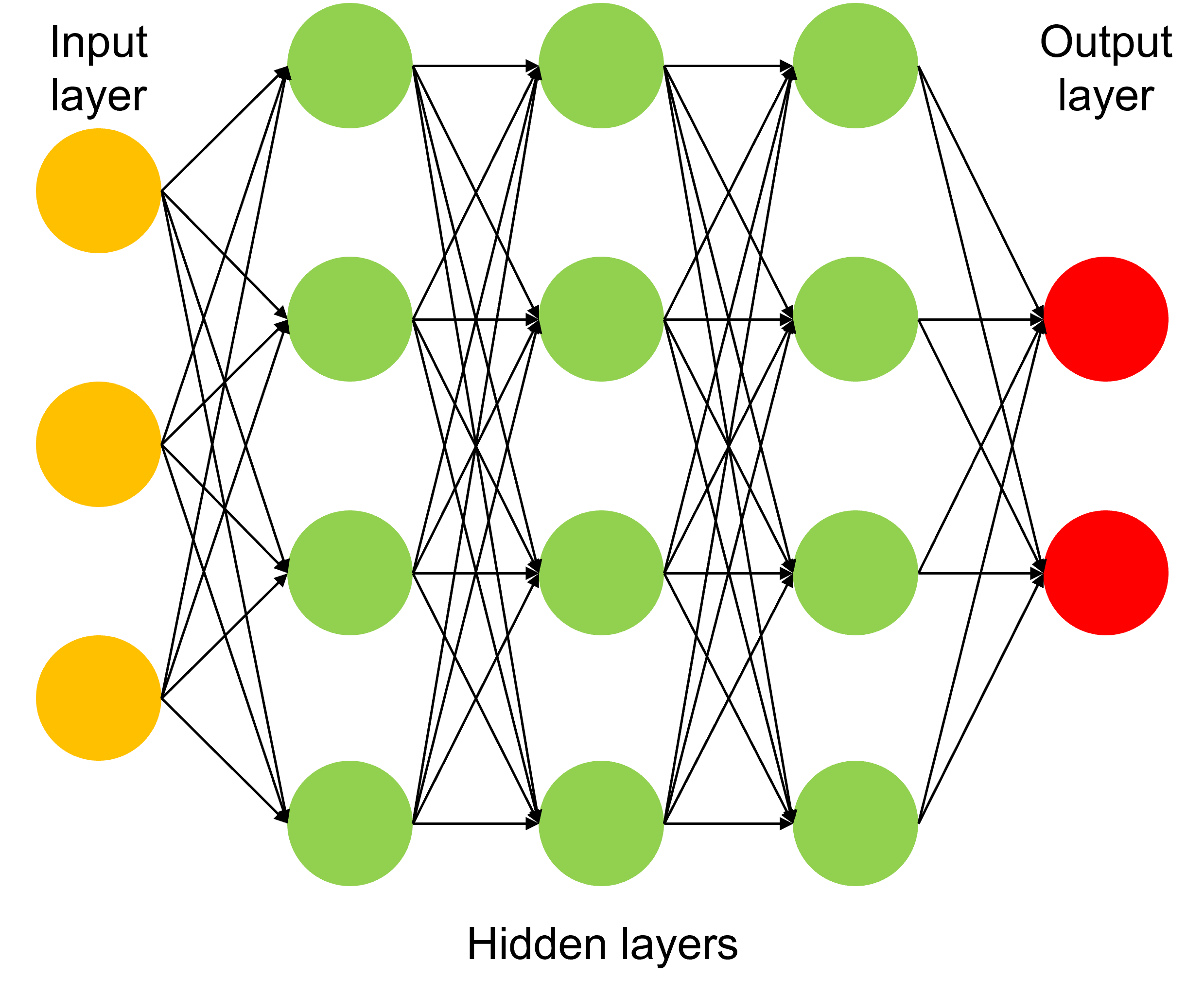}}
\caption{
Schematic representation of an artificial neural network.
}
\label{fig:ANN}
\end{center}
\end{figure}

Now as we described the mechanistic modelling workflow to evaluate the integral absolute rock permeability in the area surrounding a vertical well and the skin factor, we consider a machine learning algorithm (namely, artificial neural network or ANN) to develop the surrogate model. In Figure~\ref{fig:ANN} we show its schematic representation.  

In the framework of current study, ANN solves the regression problem. As an input, it takes the vector of 14 components, Eq.~\eqref{eq:params_ml_model}, describing the absolute permeability field $k(x, y)$ around the well inside the square of 500 m size and predicts the integral permeability $\widetilde{k}$ and skin factor $\widetilde{S}$. ANN includes input and output layers as well as several hidden layers. Each layer consists of nodes, and each node contains a number. In the first layer, the quantity of nodes is equal to the number of the input features (14 nodes in our case). In the last layer, the quantity of nodes corresponds to the number of output features (two parameters in our case as described above). The quantity of hidden layers and corresponding number of nodes are hyperparameters, which are found by conducting a series of numerical experiments. In the current machine learning model, we take 3 hidden layers with 64, 128, and 64 nodes, respectively. All ANN layers are fully-connected, so that the node $k$ in the layer $n$ is linked to all nodes in the subsequent layer $n+1$.

ANN estimates the output parameters via the forward propagation. In this procedure, the machine learning algorithm computes the values at each node, e.g., $y_{n+1, k}$, where $n+1$ is the layer number and $k$ in the node number, according to the following steps: (i) calculation of a linear combination of values at the nodes in the layer $n$ ($\mathbf{y}_n$) with weights $W_{n \rightarrow n+1, k}$; (ii) application of a non-linear activation function $g(\cdot)$ to this linear combination. Steps (i) and (ii) can be summarized as $\mathbf{y}_{n+1} = g(W_{n \rightarrow n+1} \mathbf{y}_{n} + \mathbf{b})$, where $\mathbf{b}$ denotes the bias vector. In the present model, we utilize ReLu activation function, $g(x) = \max(0, x)$. Non-linearity is used to propagate the parameters in between all layers except for the connection between the last hidden and output layers.

The weights of ANN, which can be represented as components of matrices $W_{n \rightarrow n+1}$, are tuned via the gradient descent optimization algorithm minimizing the loss function value and, in the regression task, it is a mean square error (MSE): 
\begin{equation*}
    \mathcal{L}(y, \hat{y}) = \frac{1}{n_\text{s} \cdot n_\text{out}} \sum_{i=1}^{n_\text{s}} \sum_{j=1}^{n_\text{out}} (y_{i, j} - \hat{y}_{i, j})^2,
\end{equation*}
where $y_i$ and $\hat{y}_i$ denote true and predicted output vectors for sample $i$, respectively; $n_\text{out} = 2$ is the number of output features; $n_\text{s}$ is the number of samples in a batch (mini-batch gradient descent optimization is considered) and is determined experimentally, it equals 32 in the present model. Computation of loss function gradients with respect to weights is the backward propagation. We utilize Adam optimizer \citep{kingma2014adam} as the gradient descent realization.

We apply ANN implemented in Scikit-learn library \citep{scikit-learn}, namely, function \textit{MLPRegressor()}. Synthetic dataset consisting of 40000 data points is divided into training and test sets in the ratio of 4 to 1, respectively. We tune hyperparameters of ANN including batch size, initial learning rate and strength of the L2 regularization term based on the training dataset using the random search algorithm and cross-validation technique. Random search selects randomly hyperparameter values within the predefined intervals and estimates the performance of machine learning model at a specified set of hyperparameters using the cross-validation approach. This procedure repeats 1000 times leading to the optimal combination of hyperparameters corresponding to the best model performance. We evaluate the ANN performance in terms of the mean absolute error (MAE). In the cross-validation procedure, the training dataset is divided into $M$ equal parts, $M-1$ partitions are used for training the machine learning algorithm, while the remaining part is utilized for its validation. The procedure is repeated $M$ times, and different validation part is taken at each iteration. As a result, we obtain $M$ validation (MAE) scores and average them, and this averaged metric is utilized by the random search algorithm to identify the optimal values of hyperparameters.

The random search algorithm combined with the cross-validation approach provides the following values of hyperparameters: batch size is 32, initial learning rate is $2 \cdot 10^{-3}$, prefactor before regularization term is 0.6. It is recommended to scale input features before applying ANN, which is carried out using \textit{StandardScaler()} function with parameters determined using the training data. Additionally, early stopping technique is utilized to prevent ANN from overfitting (\textit{MLPRegressor()} includes this option and leaves 10\% of the training dataset as a validation part to control the predictive capability of ANN during training phase). When the hyperparameters are tuned, ANN is trained at the training dataset, and its overall performance in terms of MAE, MSE and coefficient of determination (R$^2$) is estimated using the test set.

\begin{figure}[pos=htp]
\begin{center}
\centerline{\includegraphics[width=1\linewidth]{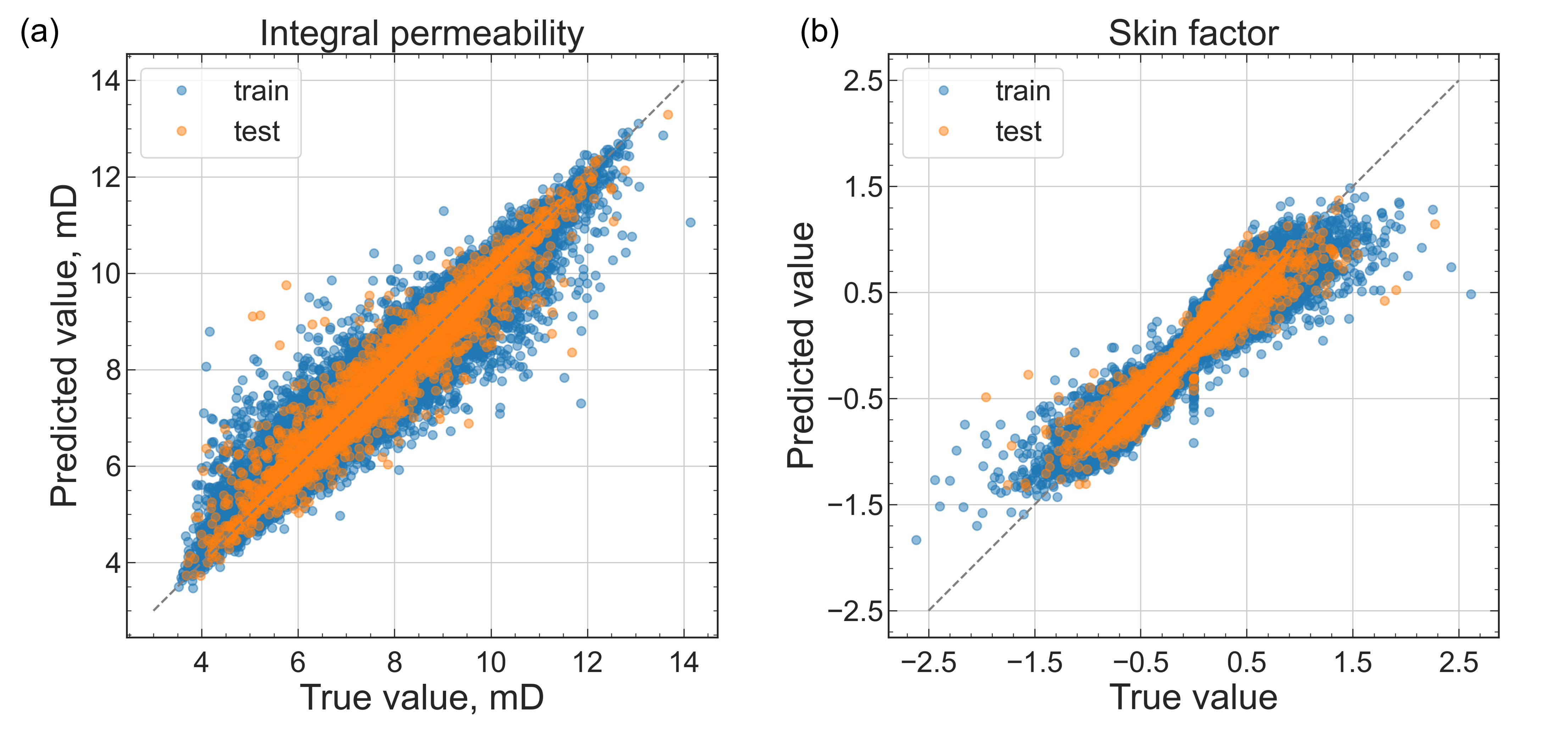}}
\caption{
Cross-plot with the predictions of the integral permeability (a) and skin-factor (b) by ANN; blue points are related to the train set, while the orange ones correspond to the test part. \label{fig:ANN_performance}}
\end{center}
\end{figure}

Figure \ref{fig:ANN_performance} shows the cross-plot with predictions of the integral permeability (a) and skin factor (b) using the surrogate model based on ANN machine learning algorithm. In this chart, the true values of output parameter are plotted at the $x$-axis, while the predicted ones are at the $y$-axis. Note that the cloud of points is oriented along the line of ideal prediction $y=x$. Only few data points are poorly estimated (with an error larger than 15\%), e.g., the samples with skin factor close to the upper and lower limits and in the vicinity of zero. We compute MAE, MSE and R$^2$ scores using the predicted and true values of the integral permeability and skin factor for the training and test datasets separately. The results are summarized in Table~\ref{tab:ANN_metrics}. We obtained the acceptable accuracy of the surrogate model based on ANN, so that it can be used for calculations of the absolute permeability map according to the methodology outlined in Section~\ref{sec:modelling_approach}.

\begin{table}[]
\centering
\caption{Surrogate model performance at train and test sets for the integral permeability and skin factor in terms of MAE, MSE and R$^2$ metrics.}
\label{tab:ANN_metrics}
\begin{tabular}{|c|c|c|c|c|}
\hline
Output parameter                       & Dataset & MAE   & MSE   & R$^2$ \\ \hline
\multirow{2}{*}{integral permeability} & train   & 0.25  & 0.153 & 0.931 \\ \cline{2-5} 
                                       & test    & 0.256 & 0.153 & 0.925 \\ \hline
\multirow{2}{*}{skin factor}           & train   & 0.08  & 0.016 & 0.909 \\ \cline{2-5} 
                                       & test    & 0.081 & 0.016 & 0.901 \\ \hline
\end{tabular}
\end{table}

\section{Results and discussion}
\label{sec:results}
\begin{figure}[pos=htp]
\begin{center}
\centerline{\includegraphics[width=1\linewidth]{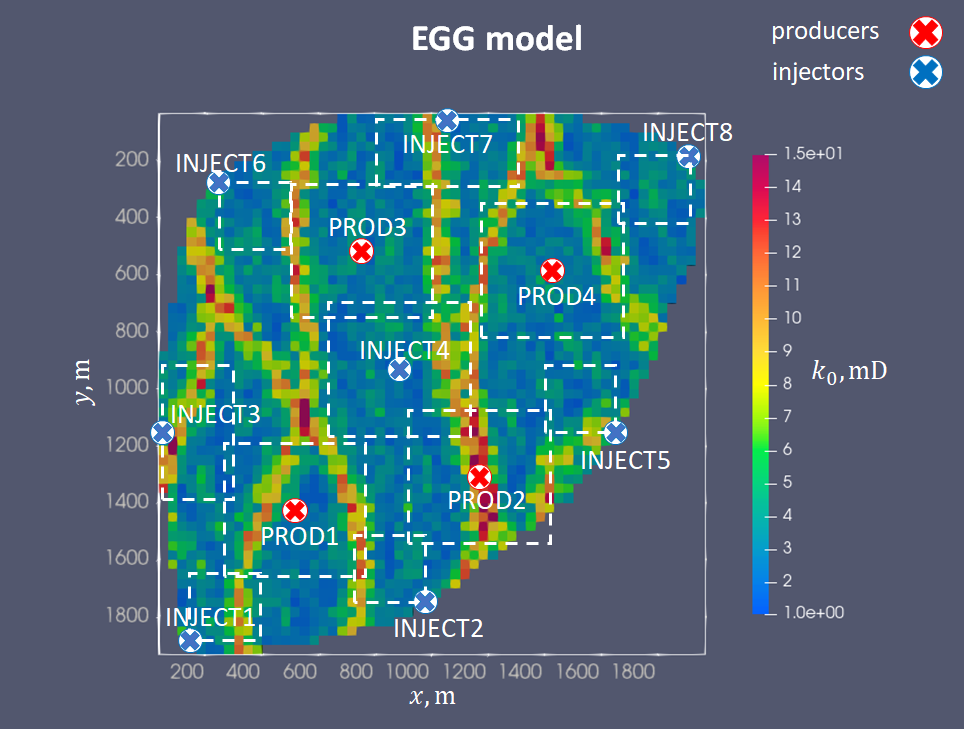}}
\caption{The top view permeability map of reservoir in ``Egg Model'' \citep{jansen2014egg}; locations of producers (red) and injectors (blue) are shown by crosses; colour map corresponds to the original permeability distribution; dashed boxes denote reservoir domains considered in physics-based averaging of the actual $k_0(x, y)$ and approximate $k(x, y)$ absolute permeability distribution around each well.}
\label{fig:EGG_model_view}
\end{center}
\end{figure}

In this section we present the results of simulations using the proposed combined mechanistic and machine learning workflow. We apply the technique outlined in Section \ref{sec:modelling_approach} to approximate the absolute permeability map of synthetic reservoir model, namely, ``Egg Model'' as described by \citet{jansen2014egg}. In Fig.~\ref{fig:EGG_model_view} we show the permeability map at the top view of the model.   

The Egg Model consists of an ensemble of 101 three-dimensional absolute permeability field realizations of a channelized oil formation. Permeability field is given in the discrete form modelled with $60\times60\times 7 = 25.000$ grid cells. The number of active blocks is 18553, and non-active cells are located around the reservoir so that its shape resembles the form of an egg. The channels have high permeability values, while the domains between them are low-permeable. Such structure of the absolute permeability field resembles the winding river patterns observed in the fluvial systems. In the current analysis, we utilize a single realization of the absolute permeability field and take its first layer. Therefore, 3600 original grid cells are considered and passed to MUFITS simulator so that our hydrodynamic model is effectively two-dimensional. The maximum and minimum values of the chosen permeability map are scaled to the interval of $[1, 15]$ mD applied for the generation of the synthetic dataset (Section~\ref{sec:surrogate_model}, equation \eqref{eq:value_ranges}), and the obtained permeability field is shown in Figure~\ref{fig:EGG_model_view}.

Let us discuss the geometrical parameters of the formation, fluid and rock properties, initial and boundary (internal and external) conditions incorporated into the Egg Model. Majority of the model parameters are similar to that described in \citep{jansen2014egg}. Numerical simulations of the two-phase (oil-water) filtration is performed using MUFITS reservoir simulator. We increase the original lateral reservoir size up to 2 km, and its thickness is set to 10 m (a single layer of mesh cells). As a result, the reservoir dimensions are 2 km $\times$ 2 km $\times$ 10 m, so that the grid block length and width are 33.3 m, while the cell height is 10 m. Porosity distribution is uniform. Oil and water are slightly compressible liquids with fixed viscosities (BLACKOIL module of MUFITS simulator is applied), while the rock is incompressible. Relative permeabilities of oil and water are governed by Corey model, and the capillary pressure is absent. At the initial state, pore pressure is uniform and equals 400 bar (the formation is located at 4 km depth), and water saturation is non-zero. The formation is intersected by 12 vertical wells, 4 producers and 8 injectors, and their locations are marked by red and blue crosses in Figure~\ref{fig:EGG_model_view} and shown in Table~\ref{tab:Egg_model_wells}. Water flooding is the major production mechanism. Vertical wells operate under constant bottomhole pressure (internal boundary condition), namely, 350 bar for producers and 450 bar for injectors, respectively. The simulation period is 30 years. All external boundaries of the reservoir are closed. Values of model parameters are summarized in Table~\ref{tab:Egg_model_params}.    

\begin{table}[pos=htp]
\centering
\caption{Parameters of the synthetic reservoir Egg Model.}
\label{tab:Egg_model_params}

\begin{tabular}{|c|c|}
\hline
Parameter                              & Value                \\ \hline
Grid-block size                               & 33.3 m $\times$ 33.3 m $\times$ 10 m                  \\ \hline
Porosity                               & 0.2                  \\ \hline
Oil compressibility                    & 10$^{-5}$ bar$^{-1}$ \\ \hline
Water compressibility                  & 10$^{-5}$ bar$^{-1}$ \\ \hline
Oil dynamic viscosity                  & 5 cP                 \\ \hline
Water dynamic viscosity                & 1 cP                 \\ \hline
End-point relative permeability, oil   & 1                 \\ \hline
End-point relative permeability, water & 1                   \\ \hline
Corey exponent, oil                    & 1                    \\ \hline
Corey exponent, water                  & 1                    \\ \hline
Residual-oil saturation                & 0.1                    \\ \hline
Connate-water saturation               & 0.1                   \\ \hline
Initial reservoir pressure             & 400 bar              \\ \hline
Initial water saturation               & 0.1                 \\ \hline
Production well bottom-hole pressures  & 350 bar              \\ \hline
Injection well bottom-hole pressures   & 700 bar              \\ \hline
Well-bore radius                       & 0.1 m                \\ \hline
Simulation time                        & 30 years             \\ \hline
\end{tabular}
\end{table}

Further, we discuss the generation of actual absolute permeability distributions $k_0(x, y)$ required for the construction the approximate fields $k(x, y)$, i.e., $k^{\mathrm{WL}}, k^{\mathrm{WT}}, S$, using the proposed method described in Section~\ref{sec:modelling_approach}. The values $\left\{k^{\mathrm{WL}}_i\right\}_{i=1}^{12}$ correspond to the absolute permeability at the well locations $\left\{x_i, y_i\right\}_{i=1}^{12}$, $k^{\mathrm{WL}}_i = k(x_i, y_i)$, and are listed in Table~\ref{tab:Egg_model_wells}. The integral permeability and skin factor are computed using the synthetic well test procedure described in Section~\ref{sec:surrogate_model}. To a well with index $j$, the following procedures are applied: (i) we cut a square of size $\mathcal{D} = 500 ~ \mathrm{m}$ with sides parallel to the coordinate axes (region $\mathcal{B}_j^{\mathcal{D}}$); (ii) pass the absolute permeability field $k_0(x, y), (x, y) \in \mathcal{B}_j^{\mathcal{D}}$ into MUFITS simulator and carry out the calculations of drawdown test (single-phase fluid problem, production with constant flow rate); (iii) conduct interpretation of the obtained bottomhole pressure behaviour using the semi-analytical reservoir model via the solution of the minimization task \eqref{eq:minimization_2}. We apply the same hydrodynamic model parameters as described in Section~\ref{sec:surrogate_model} during the synthetic well test procedure except for the flow rate and production period that, which are set to $q = 10$ m$^3$/d and $t \in [0, 360]$ d, respectively. 

\begin{figure}[pos=htp]
\begin{center}
\centerline{\includegraphics[width=1\linewidth]{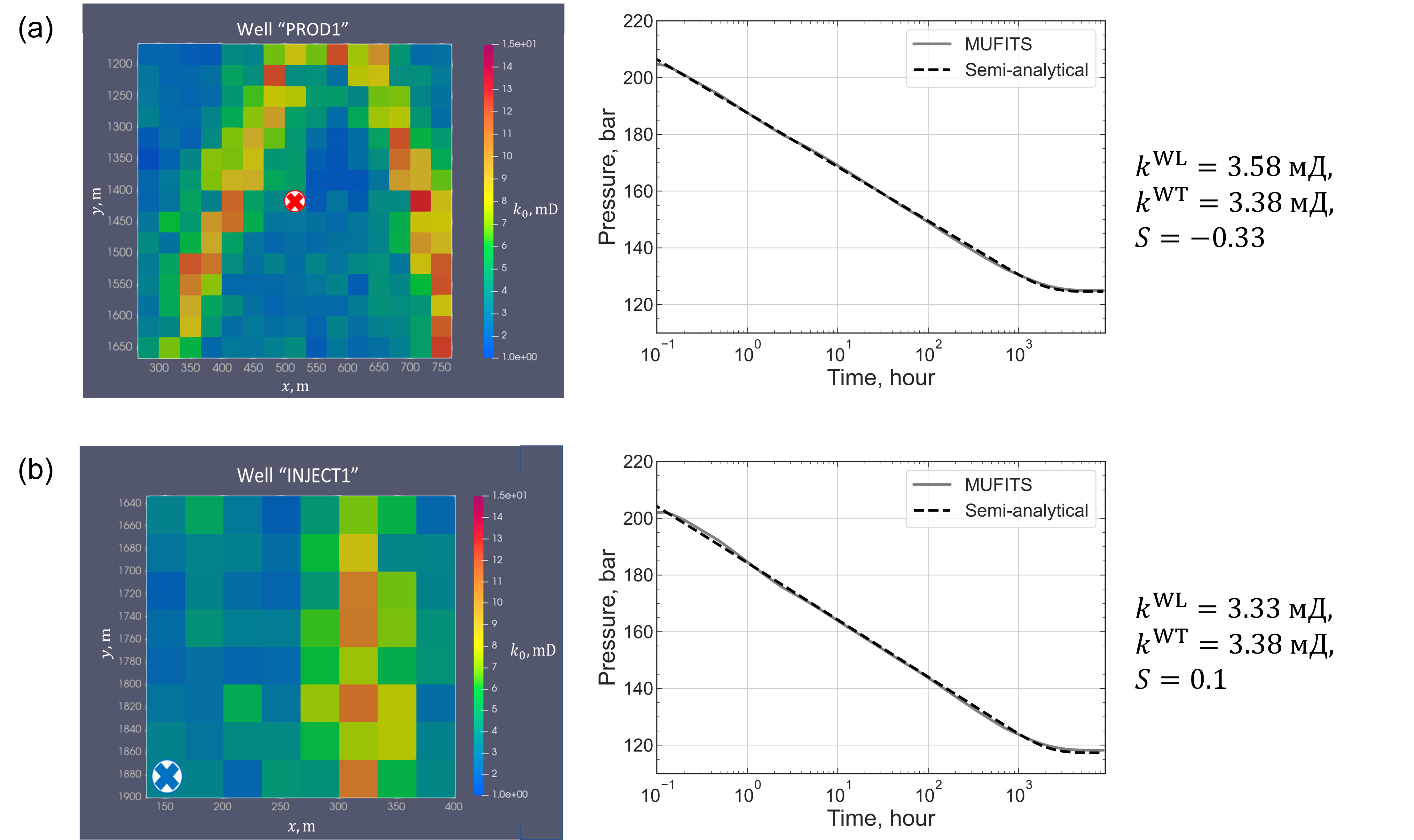}}
\caption{Results of synthetic well test procedure applied to producer ``PROD 1'' (a) and injector ``INJECT1'' (b); in the left column of plot plots the actual permeability map $k_0(x, y), (x, y) \in \mathcal{B}^{\mathcal{D}}$ is shown, while the results of the numerical drawdown test and its interpretation are given in the right column of plots; since the majority of injectors are located along the reservoir boundary, the synthetic well test is performed using the symmetry element as shown in plot (b).\label{fig:EGG_model_synth_well_test}}
\end{center}
\end{figure}

Results of synthetic well test procedure according to steps (i) -- (iii) described above and applied to the production ``PROD1'' and injection ``INJECT1'' wells are shown in Fig.~\ref{fig:EGG_model_synth_well_test}. When the absolute permeability field is essentially heterogeneous (e.g., the domain $\mathcal{B}^{\mathcal{D}}$ around the producer ``PROD2'', which is intersected by the highly permeable channel in vertical direction), the solution to the minimization problem \eqref{eq:minimization_2} providing the excellent match between the numerical bottomhole pressure dynamics and semi-analytical one does not exist and we found approximate values of $k^{\mathrm{WT}}$ and $S$. Note that it is impossible to clip a square around injectors located along the lateral border of the model, and, in this case, we perform the synthetic well test working with a symmetry element (quarter or half of the formation as shown in Fig.~\ref{fig:EGG_model_view}) in MUFITS simulator, example is shown in Figure~\ref{fig:EGG_model_synth_well_test}b. Estimated values of the integral permeability $k^{\mathrm{WT}}$ and skin factor $S$ are summarized in Table~\ref{tab:Egg_model_wells}. 
 
\begin{table}[pos=htp]
\centering
\caption{Wells locations $\left\{x_i, y_i\right\}_{i=1}^{12}$, permeability values obtained from well logging $\left\{k^{\mathrm{WL}}_i\right\}_{i=1}^{12}$ and well test $\left\{k^{\mathrm{WT}}_i\right\}_{i=1}^{12}$ (or integral permeability) measurements as well as skin factor values $\left\{S_i\right\}_{i=1}^{12}$ in the Egg Model.}
\label{tab:Egg_model_wells}
\begin{tabular}{|c|c|c|c|c|c|c|}
\hline
Index & Well ID & x, m & y, m & $k^{\mathrm{WL}}$, mD & $k^{\mathrm{WT}}$, mD & $S$     \\ \hline
1 & PROD1   & 517  & 1417 & 3.58              & 3.38                                                & -0.33 \\ \hline
2 & PROD2   & 1150 & 1317 & 11.44             & 7.92                                                & -1.35 \\ \hline
3 & PROD3   & 750  & 517  & 2.10              & 2.39                                                & 0.27  \\ \hline
4 & PROD4   & 1383 & 583  & 3.10              & 3.08                                                & -0.19 \\ \hline
5 & INJECT1 & 150  & 1883 & 3.33              & 3.38                                                & 0.10  \\ \hline
6 & INJECT2 & 983  & 1750 & 5.19              & 3.72                                                & -1.31 \\ \hline
7 & INJECT3 & 50   & 1150 & 4.22              & 5.00                                                & 0.45  \\ \hline
8 & INJECT4 & 883  & 950  & 2.00              & 2.22                                                & 0.17  \\ \hline
9 & INJECT5 & 1650 & 1150 & 6.51              & 5.14                                                & -1.01 \\ \hline
10 & INJECT6 & 250  & 283  & 3.41              & 3.22                                                & -0.24 \\ \hline
11 & INJECT7 & 1050 & 50   & 1.83              & 2.22                                                & 0.95  \\ \hline
12 & INJECT8 & 1883 & 183  & 3.68              & 3.43                                                & -0.30 \\ \hline
\end{tabular}
\end{table}

Using well locations $\left\{x_i, y_i\right\}_{i=1}^{12}$, absolute permeabilities $\left\{k^{\mathrm{WL}}_i\right\}_{i=1}^{12}$, $\left\{k^{\mathrm{WT}}_i\right\}_{i=1}^{12}$, skin factor values $\left\{S\right\}_{i=1}^{12}$ and ANN-based surrogate model described in Section~\ref{sec:surrogate_model}, we solve the optimization problem \eqref{eq:minimization_1} and estimate the values of kernel regression parameters $\Omega = \left\{\left\{k^{\mathrm{near}}_i, k^{\mathrm{far}}_i\right\}\big|_{i=1}^N, r_d, r_g, \alpha, \beta, \gamma, \delta \right\}$. Note that the solution of the optimization task \eqref{eq:minimization_1} is not unique, so that below we show a permeability map $k(x, y)$ approximating the actual distribution $k_0(x, y)$ and providing an acceptable match in terms of the oil production and water injection profiles obtained during the numerical simulations using these maps.

\begin{figure}[pos=htp]
\begin{center}
\centerline{\includegraphics[width=1\linewidth]{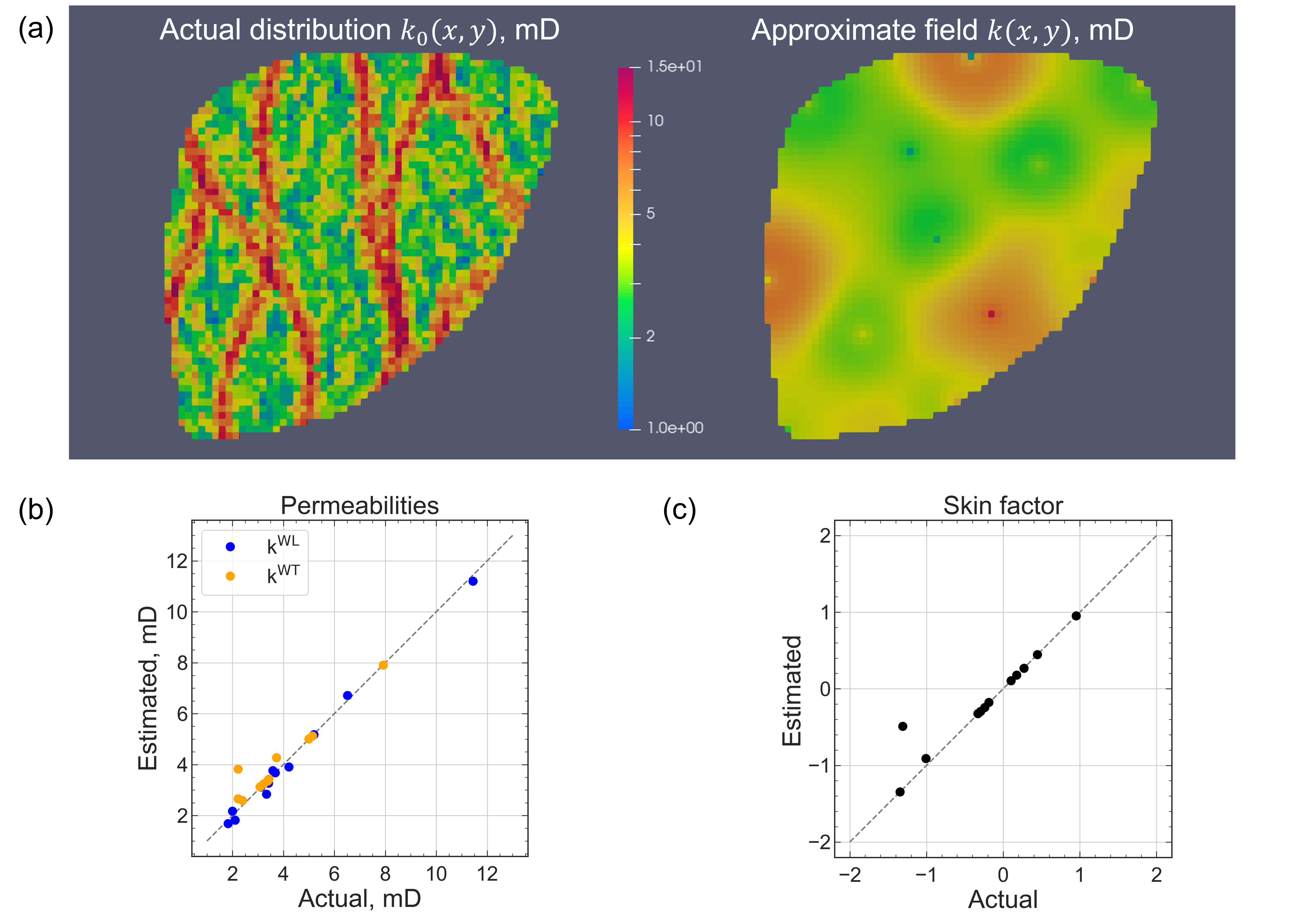}}
\caption{
Plot (a) shows the comparison of actual absolute permeability map $k_0(x, y)$ and its approximation $k(x, y)$ evaluated via equation \eqref{eq:kernel_regression} with substituted parameters from Table~\ref{tab:Egg_model_wells} and variables obtained via the solution of minimization problem \eqref{eq:minimization_1}; cross-plots (b) and (c) show actual values of $k^{\mathrm{WL}}$, $k^{\mathrm{WT}}$, $S$ and the ones obtained using approximate absolute permeability map $k(x, y)$. 
}
\label{fig:Results_map_cross_plot}
\end{center}
\end{figure}

High-permeability channels contain producer ``PROD2`` and injectors ``INJECT3'', ``INJECT7'' so that corresponding reservoir domains in the constructed approximate map have larger permeability (about 8 mD) as compared to the remaining zones, where the absolute permeability varies in between 3 and 5 mD (see Fig.~\ref{fig:Results_map_cross_plot}a). Using the approximate permeability distribution $k(x, y)$, we can determine the permeability values at the well locations $\left\{k(x_i, y_i)\right\}_{i=1}^{12}$, and, using the surrogate model, we estimate the integral permeability $\left\{\widetilde{k}_i\right\}_{i=1}^{12}$ and skin factor $\left\{\widetilde{S}_i\right\}_{i=1}^{12}$. The predicted and true values of the properties $k^{\mathrm{WL}}$, $k^{\mathrm{WT}}$, $S$ are shown in the cross-plots in Figure~\ref{fig:Results_map_cross_plot}, plots (b) and (c). Note that the integral permeability corresponding to reservoir area surrounding injector ``INJECT2'' and skin factor for the producer ``PROD2'' are estimated with large error, while the remaining values are fitted with acceptable accuracy.              

\begin{figure}[pos=htp]
\begin{center}
\centerline{\includegraphics[width=1\linewidth]{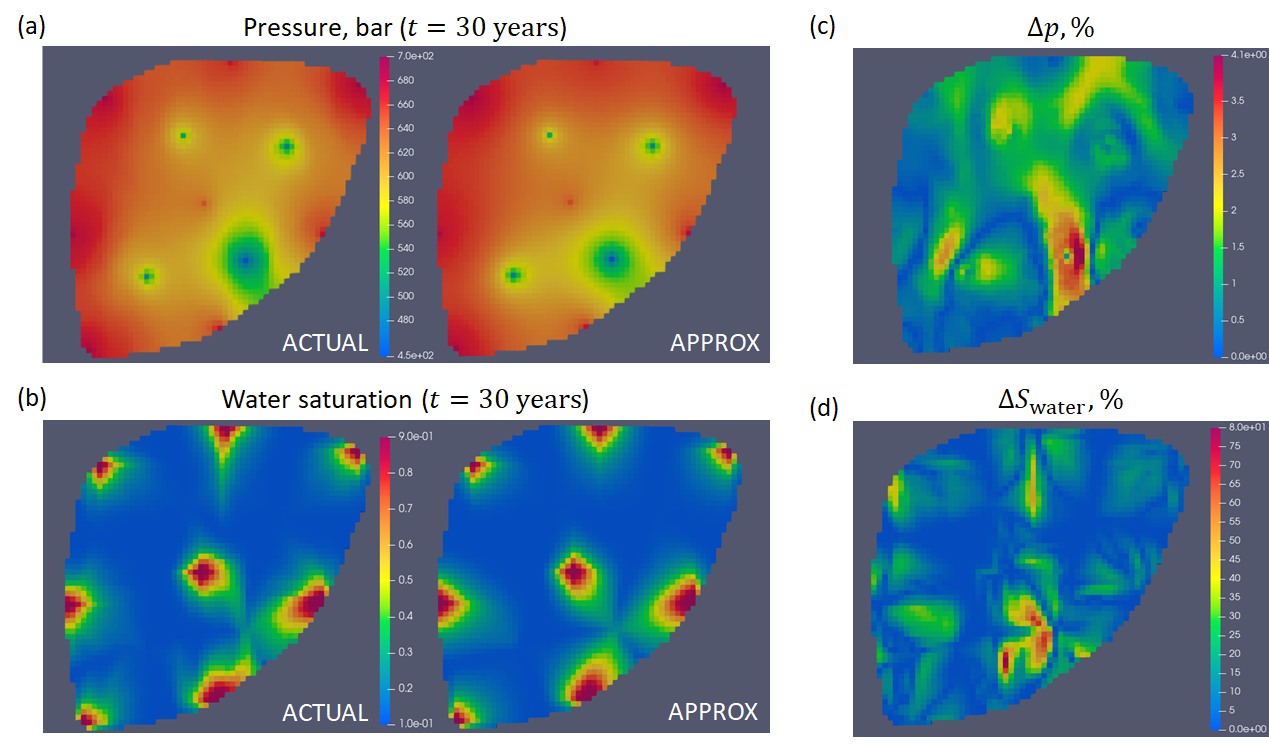}}
\caption{Plots (a) and (b) show distributions of pressure and water saturation at the end of simulation period (30 years) computed via the hydrodynamic simulator MUFITS in which the actual (ACTUAL) and approximate (APPROX) absolute permeability distributions together with the parameters outlined in Table \ref{tab:Egg_model_params} are incorporated. The relative differences between the compared distributions are shown in panels (c) and (d) for pressure $\Delta p = \left|p_{\mathrm{ACTUAL}} - p_{\mathrm{APPROX}}\right|/p_{\mathrm{ACTUAL}}$ and water saturation $\Delta S_{\mathrm{water}} = \left|(S_{\mathrm{water}})_{\mathrm{ACTUAL}} - (S_{\mathrm{water}})_{\mathrm{APPROX}}\right|/(S_{\mathrm{water}})_{\mathrm{ACTUAL}}$ fields, respectively.}
\label{fig:Results_pressure_saturation}
\end{center}
\end{figure}

Next, we carry out numerical reservoir modeling using MUFITS simulator with the parameters listed in Table~\ref{tab:Egg_model_params} combined with actual and approximate absolute permeability distributions, and we denote these cases as ``ACTUAL'' and ``APPROX'' for brevity. Figures~\ref{fig:Results_pressure_saturation}(a) and (b) show pore pressure and water saturation maps at the end of computation period (30 years), while the relative difference between the compared cases are given in plots (c) and (d). The approximate absolute permeability map allows to obtain qualitatively similar distributions of pressure and water saturation upon a fairly long operation period. Quantitative differences are small in terms of pressure (less than 4\%). As expected, relative difference in saturations $\Delta S_{\mathrm{water}}$ reaches large values in the high-permeable channels (e.g., for the domain around the producer ``PROD2'' it is 80\%), since the approximate permeability distribution can not reproduce them due to simple parametrization of permeability maps using kernel functions in the form of exponents depending on the distance to the wells. However, in low-permeability zones, the differences in the water saturation fields between analyzed solutions are small. 

\begin{figure}[pos=htp]
\begin{center}
\centerline{\includegraphics[width=1\linewidth]{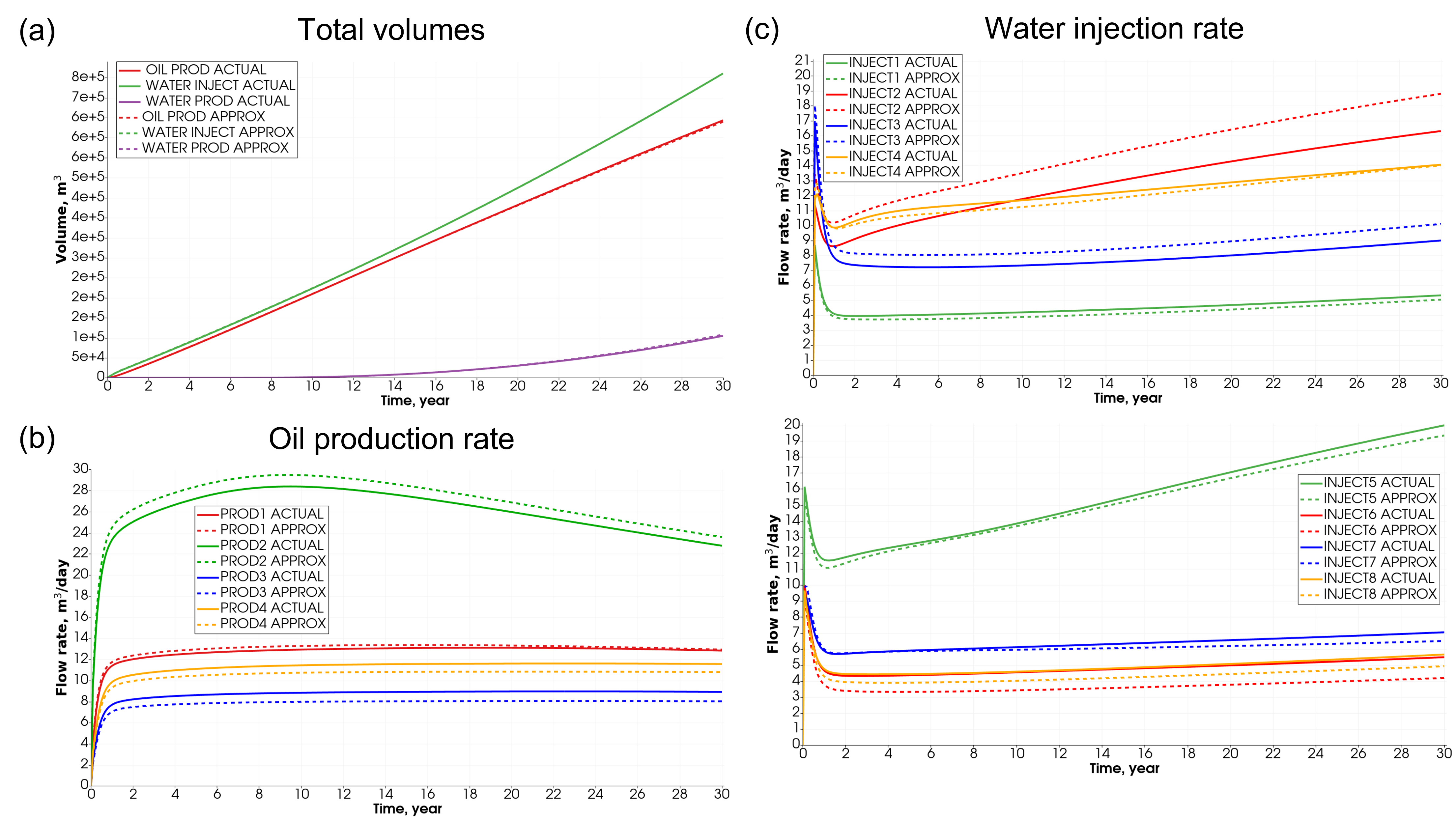}}
\caption{Plot (a) shows the dynamics of the cumulative oil (red lines) and water (green lines) production, as well as total injected water volume (magenta lines); dynamics of the oil production rate of producers ``PROD1'' -- ``PROD4'' are shown in plot (b), while plot (c) shows water injection rate histories of injectors ``INJECT1'' -- ``INJECT8''; solid lines refer to the results of numerical simulations using original permeability distribution (ACTUAL), while dashed lines corresponds to that carried out using the approximate permeability map (APPROX). \label{fig:Results_production_injection_profiles}}
\end{center}
\end{figure}

Now we analyze the temporal dependence of well flow rates. Fig.~\ref{fig:Results_production_injection_profiles}(a) shows the dynamics of cumulative production of oil and water, as well as total injected water volume. We obtained a good match between actual permeability map (solid lines) and approximate permeability map (dashed) cases. We also demonstrate oil production rate and water injection rate for all producers and injectors in plots (b) and (c) of Fig.~\ref{fig:Results_production_injection_profiles}. A notable  discrepancy between the results of reservoir simulations using original and approximated permeability maps is obtained only for injector ``INJECT2''.

Based on the obtained results shown in Figs.~\ref{fig:Results_pressure_saturation} and \ref{fig:Results_production_injection_profiles}, we conclude that the approximate permeability field $k(x, y)$ (see Fig.~\ref{fig:Results_map_cross_plot}a) is hydrodynamically similar to the actual distribution $k_0(x, y)$ since it provides the production/injection profiles close to that obtained using the original permeability map. 

\section{Discussion}
\label{sec:discussion}
In the framework of two-dimensional reservoir model and Nadaraya-Watson (NW) kernel regression we developed computationally efficient surrogate model for evaluation of integral permeability of reservoir around vertical wells as well as skin factor. As shown in the previous section, the developed model after incorporation into algorithm for construction of global absolute reservoir permeability map, allows us to obtain permeability distribution, which is hydrodynamically similar to the original one. We stress that the value of the obtained result is justified by essentially non-homogeneous permeability distribution and two-phase filtration in the synthetic reservoir, so that the considered synthetic case is close to real oilfield conditions.

Nevertheless, there are several limitations of the proposed general workflow and surrogate model of integral permeability evaluation which we would like to discuss.
\begin{enumerate}
    \item Relatively simple parameterization of permeability maps in the form of NW kernel regression based on exponent functions depending on the distance to wells. This form allows developing computationally efficient permeability cube construction algorithm, while, as a result, we obtain relatively smooth distributions, which does not allow resolving abrupt permeability variations typical of real geological conditions. In particular, failure to reproduce high-conductivity channels in fluvial geological conditions can result in significant errors in predicting water breakthrough in oilfields with water injectors;
    \item Two-dimensional problem formulation, which does not allow one to resolve heterogeneity of rock properties in vertical direction and layered reservoir structure;
    \item Missing hydraulic fractures in the problem formulation, in particular, in semi-analytical model of reservoir, as well as horizontal well trajectory (i.e. multifractured wells); as the semi-analytical reservoir model does not take into account these completions, the resulting integral permeability can be determined with a significant error, which in turn will spoil the hydrodynamic similarity and lead to poor prediction of production/injection rates in reservoir simulations using the constructed permeability maps.
\end{enumerate}

The issue with simplified spacial parametrization of permeability map can be overcome in several ways. Most obvious one is to use more sophisticated functions in kernel regression algorithm, in particular, Fourier series (or any other series of suitable basis functions) with a sufficient number of modes to resolve typical scale of heterogeneity in current geological conditions (e.g., the width of the high-permeability channels in Egg Model described above). The drawback of this approach is potentially a very large number of parameters describing permeability field (input features), which can result in very poor performance of the developed surrogate model of integral permeability evaluation. More promising direction is to develop a surrogate model to approximate integral permeability of the area surrounding a well based on a convolutional neural network (CNN) dealing with the input permeability maps in form of digital images (pictures). Input images can be grayscale, while the CNN can be either trained from a scratch or using the transfer learning approach. This approach can be combined with generative models of global reservoir permeability cube construction, in particular, generative adversarial networks. They allow creating a permeability map consistent with geological realism, which can be supplied by additional field data (e.g., by results of seismic measurement interpretations).

We believe that the second limitation described above can be overcome if a proper permeability parametrization in the vertical direction is applied, in particular, Fourier series with a number of modes sufficient to obtain a desired accuracy in current geological conditions. 

As for the third issue, we plan to develop in-house semi-analytical model to account for an effect of hydraulic fracture treatment on bottomhole pressure dynamics during welltest made in vertical fractured or horizontal multistage wells.

Application of the developed surrogate model approximating the integral absolute permeability of a certain area surrounding vertical wells is not limited by the global model of reservoir permeability cube construction as described in this study. It can be used as a computationally efficient submodel (``approximator'') evaluating the integral permeability of area surrounding wells in a wide range of global permeability cube algorithms provided the proper data flow is organized. For any permeability map generated by the global algorithm, the developed surrogate model evaluates integral permeabilities (and other integral parameters, e.g., well skin factor), which can be used to modify the parameters of global permeability distribution accordingly. 

\section{Summary and conclusions}
\label{sec:conclusions}

In this paper, we proposed a novel method to construct the approximate two-dimensional absolute permeability map $k(x, y)$. We use the available data describing the actual absolute permeability distribution $k_0(x, y)$, namely, the estimation of the absolute permeability near the well from the well-logging ($k^{\mathrm{WL}}$), the integral permeability of the zone around the well ($k^{\mathrm{WT}}$), and skin factor $S$ evaluated from the interpretation of the well test measurements. The reservoir permeability map approximation is based on Nadaraya-Watson kernel regression, and its parameters are tuned via the solution of the optimization problem, in which we minimize the differences between $k^{\mathrm{WL}}$, $k^{\mathrm{WT}}$, $S$ and their predicted values. During the solution of minimization problem, the integral permeability ($\widetilde{k}$) and skin factor ($\widetilde{S}$) around each well corresponding to the approximate permeability field are estimated by the surrogate model. It is based on an artificial neural network machine learning algorithm trained on the physics-based synthetic dataset generated with using the numerical hydrodynamic simulator (MUFITS) and in-house semi-analytical reservoir model. We applied the developed approach for the synthetic reservoir model (Egg Model). It is a two-phase oil-water problem, in which oil is produced by 4 vertical wells, while 8 injectors maintain the average pore pressure. Using the original absolute permeability distribution, we calculated the values of $k^{\mathrm{WL}}$, $k^{\mathrm{WT}}$, $S$ and constructed the approximate permeability map. By running numerical reservoir simulations using original and approximate absolute permeability maps, we obtained rather close results in terms of the well flow rates, cumulative injected/produced volumes, pressure and water saturation distributions at the end of the simulation period. As a result, the proposed approach allows to generate the permeability distribution, which is hydrodynamically similar to the actual one and eventually provides the production and injection profiles with acceptable accuracy.

The proposed surrogate model as a submodel can be incorporated in a wide range of existing methods of absolute reservoir permeability cube construction for effective well test data fusion. This can be carried out if the developed model is treated as an ``approximator'', which at the input takes the permeability map in a certain area surrounding the well and returns its integral value and well skin factor.  

There are several directions for future research related to the developed models, namely, taking into account geological realism (i.e., complex distributions of permeability either in the form of figures or using complex parametrization), generalization of the proposed method to 3D taking into account heterogeneity of the reservoir permeability in vertical direction and taking into account a vertical fractured well  or multiple fractures in horizontal wells. 

\section*{Acknowledgements}
The work was supported by the Analytical center under the RF Government (subsidy 460 agreement \\ 000000D730321P5Q0002, Grant No. 70-2021-00145 02.11.2021). 

\section*{Data availability}
The original input file for the reservoir simulator MUFITS containing the Egg Model is given in the website: \\ \href{http://mufits.org/example-egg-model.html}{The Egg Model Study}. The data files for the EGG Model can be found in the website: \href{https://data.4tu.nl/articles/dataset/The_Egg_Model_-_data_files/12707642/1}{The Egg Model - data files}.

\section*{Code availability}

Repository name: Data\_Fusion\_HDM\_ML

Program language: Python
 
Software required: Python libraries -- numpy, pandas, scipy, sklearn, math, pickle, joblib, matplotlib

Program size: 11 MB

The source codes are available for downloading at the link: \href{https://github.com/evgenii-kanin/Data_Fusion_HDM_ML}{https://github.com/evgenii-kanin/Data\_Fusion\_HDM\_ML}.


\bibliographystyle{cas-model2-names}
\bibliography{main_text}

\end{document}